\documentclass[lettersize,journal]{IEEEtran}
\usepackage{amsmath,amsfonts}
\usepackage{algorithmic}
\usepackage{algorithm}
\usepackage{array}
\usepackage{stfloats}
\usepackage{url}
\usepackage{verbatim}
\usepackage{graphicx}
\usepackage{gensymb}
\usepackage{caption}
\usepackage{subcaption}
\usepackage{multicol}
\usepackage{multirow}
\usepackage{url}
 \usepackage{hyperref}
 \hypersetup{colorlinks=false,pdfborder={0 0 0}}

\def\UnderWiggleTemp{\the\catcode`\@}
\catcode`\@=11
\ifx\UnderWiggle@Loaded\relax
  \catcode`\@=\UnderWiggleTemp
    \else \let\UnderWiggle@Loaded=\relax \fi
\newbox\U@BoxA
\newbox\U@BoxB
\newdimen\U@DimenA
\def\U@DoUnderWiggle{
  \offinterlineskip
  \vtop{
    \hbox{\vbox{\copy0}}
    \vskip 1.2pt  
    \vbox to 0.4pt{
      \hbox to\wd0{\hss\char'176\hss}
      \vskip0pt minus 1fil
    }
    \vskip 0.4pt  
  }
}
\def\UnderWiggle#1{{%
  \ifmmode
    \mathchoice
      {\setbox0=\hbox{$\displaystyle #1$}\U@DoUnderWiggle}
      {\setbox0=\hbox{$\textstyle #1$}\U@DoUnderWiggle}
      {\setbox0=\hbox{$\scriptstyle #1$}\U@DoUnderWiggle}
      {\setbox0=\hbox{$\scriptscriptstyle #1$}\U@DoUnderWiggle}
  \else
    \setbox0=\hbox{#1}\U@DoUnderWiggle
  \fi
}}
\catcode`\@=\UnderWiggleTemp

\newcommand{\be}{\begin{eqnarray}}         \newcommand{\uw}{\UnderWiggle}
\newcommand{\ee}{\end{eqnarray}}           \newcommand{\ba}{\begin{eqnarray*}}
\newcommand{\ea}{\end{eqnarray*}}

 \newcommand{\bl}{\begin{lemma}}
\newcommand{\el}{\end{lemma}}              \newcommand{\bd}{\begin{definition}}
                 \newcommand{\ed}{\end{definition}}
               
 \newcommand{\bc}{\begin{corollary}}
\newcommand{\ec}{\end{corollary}}          \newtheorem{prop}{Proposition}
\newcommand{\bp}{\begin{prop}}             \newcommand{\ep}{\end{prop}}
\newcommand{\One}{\mbox{\large \bf 1}}

\def\boldfacefake #1{%
    \hbox{%
        \mathsurround=0pt
        \hbox to 0.4pt{$#1$\hss}%
        \hbox to 0.4pt{$#1$\hss}%
        \hbox {$#1$}%
    }%
}

\newcommand{\twofslides}[4]
{
\hbox to\hsize{\hss
    \vbox{\psfig{figure=\mydir/#1,width=#3,height=#4}}\qquad
    \vbox{\psfig{figure=\mydir/#2,width=#3,height=#4}}
    \hss}
\vskip 0.0truein
}

\newcommand{\twof}[4]
{
\hbox to\hsize{\hss
    \vbox{\psfig{figure=\mydir/#1,width=#3,height=#4}}   \qquad
    \vbox{\psfig{figure=\mydir/#2,width=#3,height=#4}}
    \hss}
}
\newcommand{\twofigspec}[2]
{
\hbox to\hsize{\hss
    \vbox{\psfig{figure=\mydir/#1,width=2.7in,height=3.6in}}\qquad
    \vbox{\psfig{figure=\mydir/#2,width=2.7in,height=3.6in}}
    \hss}
\vskip -0.1truein
\hbox to\hsize{\hss
    \vbox{ \begin{center}\mbox{\footnotesize \hspace{0.0in} {(a)}
                     \hspace{2.5in} {(b)}  }  \end{center} }
    \hss}
}
\newcommand{\twofig}[2]
{
\hbox to\hsize{\hss
    \vbox{\psfig{figure=\mydir/#1,width=2.7in,height=2.7in}}\qquad
    \vbox{\psfig{figure=\mydir/#2,width=2.7in,height=2.7in}}
    \hss}
\vskip -0.1truein
\hbox to\hsize{\hss
    \vbox{ \begin{center}\mbox{\footnotesize \hspace{0.2in} {(a)}
                     \hspace{2.7in} {(b)}  }  \end{center} }
    \hss}
}
\newcommand{\twofigland}[2]
{
\hbox to\hsize{\hss
    \vbox{\psfig{figure=\mydir/#1,width=2.8in,height=1.8in}}\qquad
    \vbox{\psfig{figure=\mydir/#2,width=2.8in,height=1.8in}}
    \hss}
\vskip -0.1truein
\hbox to\hsize{\hss
    \vbox{ \begin{center}\mbox{\footnotesize \hspace{0.1in} {(a)}
                     \hspace{2.8in} {(b)}  }  \end{center} }
    \hss}
}
\newcommand{\twofigl}[2]
{
\hbox to\hsize{\hss
    \vbox{\psfig{figure=\mydir/#1,width=1.9in,height=1.9in}}\qquad
    \vbox{\psfig{figure=\mydir/#2,width=1.9in,height=1.9in}}
    \hss}
\vskip -0.1truein
\hbox to\hsize{\hss
    \vbox{ \begin{center}\mbox{\footnotesize \hspace{0.1in} {(a)}
                     \hspace{1.9in} {(b)}  }  \end{center} }
    \hss}
}
\newcommand{\twofigsq}[2]
{
\hbox to\hsize{\hss
    \vbox{\psfig{figure=\mydir/#1,width=2.3in,height=2.3in}}\qquad
    \vbox{\psfig{figure=\mydir/#2,width=2.3in,height=2.3in}}
    \hss}
\vskip -0.1truein
\hbox to\hsize{\hss
    \vbox{ \begin{center}\mbox{\footnotesize \hspace{0.1in} {(a)}
                     \hspace{2.3in} {(b)}  }  \end{center} }
    \hss}
}
\newcommand{\twofigo}[2]{
\hbox to\hsize{\hss
    \vbox{\psfig{figure=\mydir/#1,width=2.2in,height=2.6in}}\qquad
    \vbox{\psfig{figure=\mydir/#2,width=2.2in,height=2.6in}}
    \hss}
\vskip -0.1truein
\hbox to\hsize{\hss
    \vbox{ \begin{center}\mbox{\footnotesize \hspace{0.1in} {(a)}
                     \hspace{2.2in} {(b)}  }  \end{center} }
    \hss}
}
\newcommand{\twofigos}[2]{
\hbox to\hsize{\hss
    \vbox{\psfig{figure=\mydir/#1,width=2.1in,height=2.6in}}\qquad
    \vbox{\psfig{figure=\mydir/#2,width=2.1in,height=2.6in}}
    \hss}
\vskip -0.4truein
\hbox to\hsize{\hss
    \vbox{ \begin{center}\mbox{\footnotesize \hspace{0.1in} {(a)}
                     \hspace{2.1in} {(b)}  }  \end{center} }
    \hss}
}
\newcommand{\twosmall}[2]{
\hbox to\hsize{\hss
    \vbox{\psfig{figure=\mydir/#1,width=1.3in}} \hspace{0.8in}
    \vbox{\psfig{figure=\mydir/#2,width=1.3in}}
    \hss}
\vskip -0.1truein
\hbox to\hsize{\hss
    \vbox{ \begin{center}\mbox{\footnotesize \hspace{0.1in} {(a)}
                     \hspace{1.9in} {(b)}  }  \end{center} }
    \hss}
}

\newcommand{\threef}[5]
{
\hbox to\hsize{\hss
  \vbox{\psfig{figure=\mydir/#1,width=#4,height=#5}}%
  \hss}
\hbox to\hsize{\hss
  \vbox{\psfig{figure=\mydir/#2,width=#4,height=#5}}%
  \hss}
\hbox to\hsize{\hss
  \vbox{\psfig{figure=\mydir/#3,width=#4,height=#5}}%
  \hss}
}

\newcommand{\fourf}[6] 
{
\hbox to\hsize{\hss
    \vbox{\psfig{figure=\mydir/#1,width=#5,height=#6}}\qquad
    \vbox{\psfig{figure=\mydir/#2,width=#5,height=#6}}
    \hss}
\vskip 0.1truein
\hbox to\hsize{\hss
    \vbox{\psfig{figure=\mydir/#3,width=#5,height=#6}}\qquad
    \vbox{\psfig{figure=\mydir/#4,width=#5,height=#6}}
    \hss}
\vskip -0.1truein
}
\newcommand{\fourfig}[6]
{
\hbox to\hsize{\hss
    \vbox{\psfig{figure=\mydir/#1,width=#5,height=#6}}\qquad
    \vbox{\psfig{figure=\mydir/#2,width=#5,height=#6}}
    \hss}
\vskip -0.1truein
\hbox to\hsize{\hss
    \vbox{ \begin{center}\mbox{\footnotesize \hspace{0.1in} {(a)}
                     \hspace{#5} {(b)}  }  \end{center} }
    \hss}
\vskip 0.1truein
\hbox to\hsize{\hss
    \vbox{\psfig{figure=\mydir/#3,width=#5,height=#6}}\qquad
    \vbox{\psfig{figure=\mydir/#4,width=#5,height=#6}}
    \hss}
\vskip -0.1truein
    \vbox{ \begin{center}\mbox{\footnotesize \hspace{0.1in} {(c)}
                     \hspace{#5} {(d)}  }  \end{center} }
\hbox to\hsize{\hss
    \hss}
\vskip -0.1truein
}
\newcommand{\twofv}[4]
{
\hbox to\hsize{\hss
    \vbox{\psfig{figure=\mydir/#1,width=#3,height=#4}}
    \hss}
\vskip -0.1truein
\hbox to\hsize{\hss
    \vbox{\begin{center} \footnotesize{(a)} \end{center}}
    \hss}
\vskip 0.1truein
\hbox to\hsize{\hss
    \vbox{\psfig{figure=\mydir/#2,width=#3,height=#4}}
    \hss}
\vskip -0.1truein
\hbox to\hsize{\hss
    \vbox{\begin{center} \footnotesize{(b)} \end{center}}
    \hss}
\vskip -0.1truein
}
\newcommand{\sixfig}[8]
{
\hbox to\hsize{\hss
    \vbox{\psfig{figure=\mydir/#1,width=#7,height=#8}}\qquad
    \vbox{\psfig{figure=\mydir/#2,width=#7,height=#8}}
    \hss}
\vskip -0.1truein
\hbox to\hsize{\hss
    \vbox{ \begin{center}\mbox{\footnotesize {(a)}
                     \hspace{#7} {(b)}  }  \end{center} }
    \hss}
\vskip 0.1truein
\hbox to\hsize{\hss
    \vbox{\psfig{figure=\mydir/#3,width=#7,height=#8}}\qquad
    \vbox{\psfig{figure=\mydir/#4,width=#7,height=#8}}
    \hss}
\vskip -0.1truein
\hbox to\hsize{\hss
    \vbox{ \begin{center}\mbox{\footnotesize  {(c)}
                     \hspace{#7} {(d)}  }  \end{center} }
    \hss}
\vskip 0.1truein
\hbox to\hsize{\hss
    \vbox{\psfig{figure=\mydir/#5,width=#7,height=#8}}\qquad
    \vbox{\psfig{figure=\mydir/#6,width=#7,height=#8}}
    \hss}
\vskip -0.1truein
\hbox to\hsize{\hss
    \vbox{ \begin{center}\mbox{\footnotesize  {(e)}
                     \hspace{#7} {(f)}  }  \end{center} }
    \hss}
}

\newcommand{\threespec}[6]{
\hbox to\hsize{\hss
    \vbox{\psfig{figure=\mydir/#1,width=#4,height=#5}} \hspace{0.1in}
    \vbox{\psfig{figure=\mydir/ieq.eps,width=#6,height=#5}} \hspace{0.1in}
    \vbox{\psfig{figure=\mydir/#2,width=#4,height=#5}} \hspace{0.1in}
    \vbox{\psfig{figure=\mydir/iplus.eps,width=#6,height=#5}} \hspace{0.1in}
    \vbox{\psfig{figure=\mydir/#3,width=#4,height=#5}}
    \hss}
}
\newcommand{\sixf}[8]{
\hbox to\hsize{\hss
    \vbox{\psfig{figure=\mydir/#1,width=#7,height=#8}} \hspace{0.1in}
     \hspace{0.1in}
    \vbox{\psfig{figure=\mydir/#2,width=#7,height=#8}} \hspace{0.1in}
    \hspace{0.1in}
    \vbox{\psfig{figure=\mydir/#3,width=#7,height=#8}}
    \hss}
\vskip 0.1truein
    \hbox to\hsize{\hss
    \vbox{\psfig{figure=\mydir/#4,width=#7,height=#8}} \hspace{0.1in}
   \hspace{0.1in}
    \vbox{\psfig{figure=\mydir/#5,width=#7,height=#8}} \hspace{0.1in}
  \hspace{0.1in}
    \vbox{\psfig{figure=\mydir/#6,width=#7,height=#8}}
    \hss}
}

\newcommand{\onee}[3]  
{
\includegraphics[width=#2,height=#3]{\mydir/#1}
}

\newcommand{\twoo}[6]  
{
\mbox{\subfigure{\includegraphics[width=#3,height=#4]{\mydir/#1}}\quad \quad
      \subfigure{\includegraphics[width=#5,height=#6]{\mydir/#2}} }
}

\newcommand{\threee}[5]  
{
\mbox{\subfigure{\includegraphics[width=#4,height=#5]{\mydir/#1}}\quad
      \subfigure{\includegraphics[width=#4,height=#5]{\mydir/#2}}\quad
      \subfigure{\includegraphics[width=#4,height=#5]{\mydir/#3}} }}

\newcommand{\fourr}[6]  
{
\mbox{\subfigure{\includegraphics[width=#5,height=#6]{\mydir/#1}}\quad
      \subfigure{\includegraphics[width=#5,height=#6]{\mydir/#2}} }
\mbox{\subfigure{\includegraphics[width=#5,height=#6]{\mydir/#3}}\quad
      \subfigure{\includegraphics[width=#5,height=#6]{\mydir/#4}} }
}

   {\VerbatimEnvironment
    \begin{Sbox}\begin{minipage}{10cm}\begin{Verbatim}}%
   {\end{Verbatim}\end{minipage}\end{Sbox}
    \setlength{\fboxsep}{8pt}\fbox{\TheSbox}}

\usepackage{cite}

\begin{document}

\title{A Multiscale Approach for Enhancing Weak Signal Detection}

\author{Dixon Vimalajeewa$^*$, Ursula U. Müller, and Brani Vidakovic,~\IEEEmembership{Member, IEEE}}

\maketitle

\begin{abstract}

Stochastic resonance (SR), a phenomenon originally introduced in climate modeling, enhances signal detection by leveraging optimal noise levels within non-linear systems. Traditional SR techniques, mainly based on single-threshold detectors, are limited to signals whose behavior does not depend on time. Often large amounts of noise are needed to detect weak signals, which can distort complex signal characteristics. To address these limitations, this study explores multi-threshold systems and the application of SR in multiscale applications using wavelet transforms. In the multiscale domain signals can be analyzed at different levels of resolution to better understand the underlying dynamics.

We propose a double-threshold detection system that integrates two single-threshold detectors to enhance weak signal detection. We evaluate it both in the original data domain and in the multiscale domain using simulated and real-world signals and compare its performance with existing methods.

Experimental results demonstrate that, in the original data domain, the proposed double-threshold detector significantly improves weak signal detection compared to conventional single-threshold approaches. Its performance is further improved in the frequency domain, requiring lower noise levels while outperforming existing detection systems. This study advances SR-based detection methodologies by introducing a robust approach to weak signal identification, with potential applications in various disciplines. 

\end{abstract}


{
  \renewcommand{\thefootnote}{}%
  \footnotetext[1]{The author Dixon Vimalajeewa is with the Department of Statistics, University of Nebraska Lincoln, Nebraska, USA; Ursula U. Müller and Brani Vidakovic are with the Department of Statistics, Texas A\&M University, College Station, Texas, USA.}
  \footnotetext[2]{$^*$Corresponding author: himalajeewa2@unl.edu}
}

\section{Introduction}
\IEEEPARstart{T}he concept of stochastic resonance (SR) was introduced by Benzi et al. \cite{benzi1981} in climate modeling to describe the periodic recurrence of ice ages. The recurrence of ice ages can be explained by a signal, the orbital eccentricity, which alone is too weak to have an effect. It must therefore be assisted by other factors, modeled as noise, which help the signal to switch from one state point to the other. This phenomenon is called ``stochastic resonance".

Stochastic resonance can also be applied to enable the detection of signals that are typically too weak for conventional detectors, by artificially adding noise. SR occurs when the signal-to-noise ratio in a non-linear system improves by increasing the noise so that a weak signal becomes detectable. Since its introduction, the concept of stochastic resonance has attracted a lot of attention in many fields, for example in physics, neuroscience and engineering, where it has found various applications \cite{wiesenfeld1995, greenwood2000, zheng2014}.

Stochastic resonance enables signal detection from threshold data, i.e. when a noisy signal exceeds one or more thresholds, without requiring a detailed understanding of the system’s underlying dynamics. That is, thresholding creates a discontinuous, piecewise response to an input signal, so the output depends on whether the input exceeds a certain threshold value, allowing SR studies to be conducted independently of specific system characteristics \cite{Gammaitoni1998}. Threshold detectors are therefore useful for investigating SR in complex systems. Threshold data can be regarded as pulses that are generated when the input signal surpasses a predefined threshold.
At high noise levels, the probability of threshold exceedance can be quite large (up to 50\% if the noise distribution is symmetric), whereas at low noise levels it remains close to zero

\begin{figure}[!t]
    \vspace{-1.2cm}
        \centering
            \includegraphics[width= 1\columnwidth]{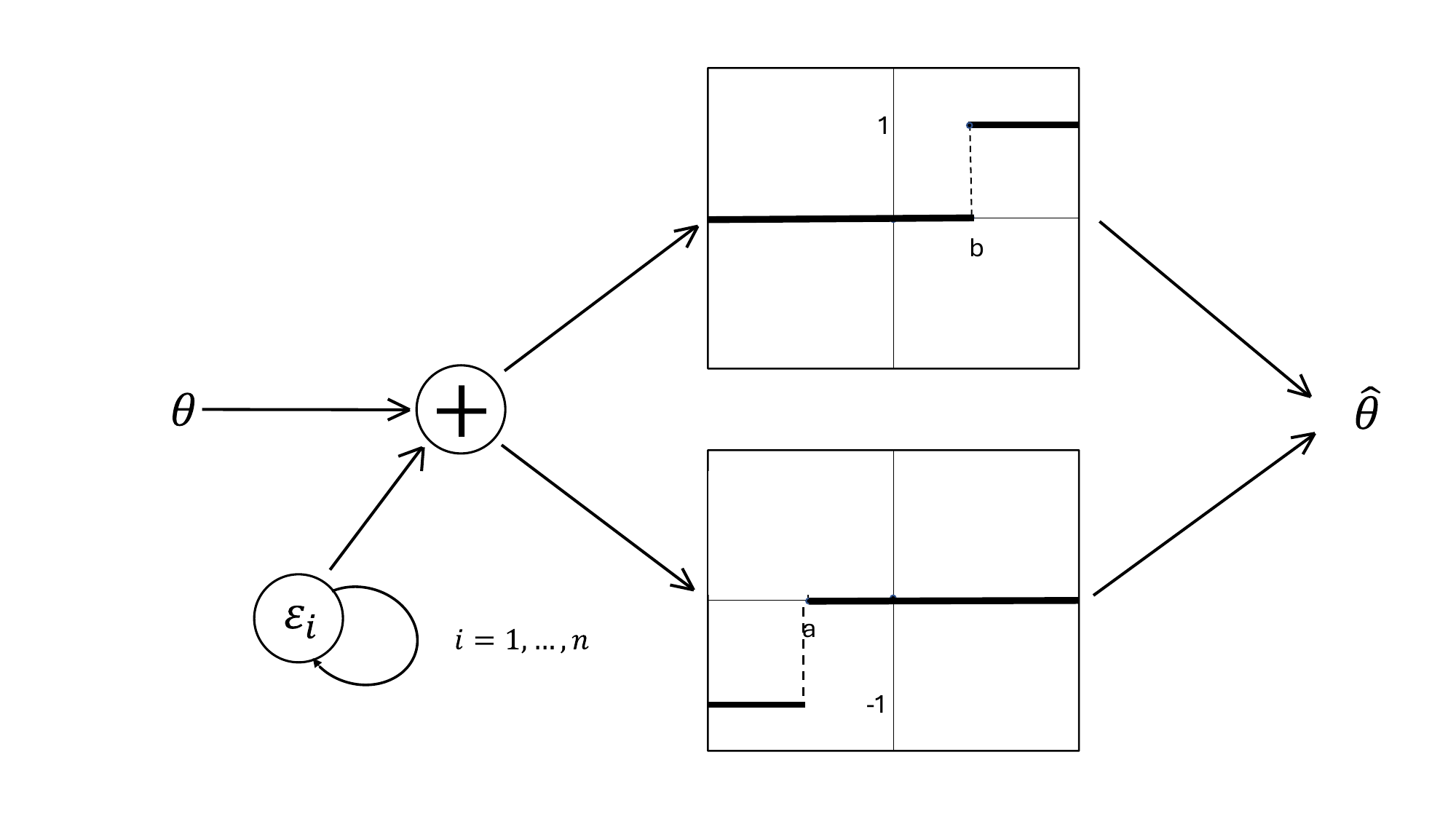}
             \caption{Stochastic resonance with threshold detector: schematic representation of a double-threshold detector and signal input $\theta$. The detector produces indicators $-1$ and $1$ in response to the events $X  < a$ and $X > b$, where $X = \theta + \epsilon_i$. These event indicators are used to compute an estimator $\hat \theta$ for the signal $\theta$}. 
    \label{fig:schema}
\end{figure}

The core principle of SR is to identify an optimal noise level referred to as the ``SR point'' \cite{greenwood2003}, at which the observed threshold exceedances most accurately reflect variations in signal intensity. Introducing this optimal noise level enables the detection of weak signals that would otherwise remain undetectable. Studies also suggest that multi-threshold systems can enhance weak signal detection. (See, e.g., Figure \ref{fig:schema}, which illustrates a double-threshold system.) Greenwood et al. \cite{greenwood1999} demonstrated for a scenario with standard normal noise, that a single-threshold system retains up to 0.6366 information, while a dual-threshold system increases this retention up to 0.7596. However, a critical challenge of the existing threshold systems is that they are predominantly limited to constant (time-invariant) signals. As a result, their applicability is limited as most of the signals monitored at real-world processes are time-variant. This in turn has increased the demand for advanced threshold detectors capable of detecting time-variant signals. 

Standard SR techniques often require a large amount of noise, so that more complex signal characteristics cannot be restored. A promising alternative is to implement SR in a different domain, where the requirement for large noise amplitudes is reduced, enabling the recovery of finer signal details. Signals in their multiscale (frequency) domain, obtained by using wavelet transforms, have proven to be effective in enhancing hidden signal properties. More specifically, 
the strength of wavelet transforms lies in their ability to decompose signals into discrete, time-localized frequency components, meaning they can precisely identify both the frequency content of a signal and the specific time intervals during which these frequencies occur. Through recursive decomposition, only low-frequency (smooth) approximation coefficients are retained, progressively refining the signal’s low-frequency structure at each level \cite{Vidakovic2009}. This process enhances the likelihood of exceeding a given threshold, particularly in higher decomposition levels. Literature, such as \cite{Bhuvaneshwari2013, Chouhan2012}, illustrates that wavelet transform-based SR systems improve contrast in dark images, outperforming conventional methods like the discrete cosine transform. The existing wavelet transform based SR systems are primarily based on constant signals such as images.

The application of wavelet-based SR methods to non-constant (time-varying) signals is little researched, despite their great potential in many practical scenarios. A detailed motivational example that explains the usefulness of wavelets in a communication scenario is provided in Section \ref{subsec:motivation}. In that scenario, a receiver observes only signal values outside a given band. The use of wavelets can be seen as a preprocessing tool that makes it possible to identify details of the signal with greater accuracy.
 
This study has two primary objectives. First, it aims to highlight the significance of employing SR techniques with multi-threshold systems for detecting time-varying signals. Second, it investigates the application of SR in the multiscale and frequency domains to enhance weak signal detection while reducing the reliance on high noise levels. A key contribution of this study is the development of a novel estimator for double-threshold systems. To assess its effectiveness, we conducted experiments evaluating its ability to detect time-varying weak signals in both the time and frequency domains. The wavelet transform is used to convert signals into the frequency domain. The experiments involve various one-dimensional (1D) and two-dimensional (2D) signals, including both simulated data and commonly used signals. The results demonstrate that the proposed double-threshold detector outperforms standard single-threshold detectors in detecting weak signals. Additionally, its performance improves further, with reduced noise, when applied in the frequency domain. These findings provide new insights into enhancing SR techniques for weak signal detection, contributing to advancements in practical signal processing applications.

This paper is structured as follows. In Section \ref{sec:Problem} we present the formulation of our problem and derive an estimator designed to detect weak signals using a double-threshold SR system. The performance evaluations are described in Section \ref{sec:Simulation} while Section \ref{sec:results} presents evaluation outcomes. In Section \ref{sec:Discussion}, we examine and discuss these results. Finally, the paper concludes with the summarizing remarks presented in Section \ref{sec:Conclusion}. 

\section{Estimation} \label{sec:Problem}
Let $a,b\in\mathbb{R}$, $a < b$, denote two fixed thresholds. 
Assume that a receiver can record and distinguish only the time points when 
a weak signal $X_i$, $i=1,\ldots n$, exceeds the upper threshold $b$ or goes 
below the lower threshold $a$. The observations can be coded with indicators,
\be \label{eq:indicators}
Y_i = \One(X_i >b) - \One(X_i <a),
\ee
$i = 1,\ldots, n$, 
i.e.\ one observes a sequence of size $n$ of zeroes and plus/minus one's.

We assume in the following that the signal is an undetectable constant 
$\theta \in (a,b)$. It is possible to make inferences about $\theta$ from the data (\ref{eq:indicators}) on the receiving end if the signal is or can be convolved with a random noise of controlled size, prior to getting to the detector (Fig. \ref{fig:schema}). Hence we consider an  SR scenario with a weak constant signal $\theta$ that is enhanced by an additive noise $\epsilon_i$, so the receiver gets the information whether the signal is above $b$, below $a$, or not visible. Formally we have
\be \label{eq:model}
X_i = \theta + \epsilon_i, \quad i = 1,\dots,n,
\ee
where the $\epsilon_i$ are independent identically distributed (i.i.d.) random variables with mean zero, variance $\sigma^2 \in (0,\infty)$, and a known invertible distribution function $F$. 


\subsection{Estimation of $\theta$ using both thresholds separately} \label{sec:thhat}
The data (\ref{eq:indicators}) are categorical. We therefore have a multinomial distribution of size $n$ with three categories, $+1$, $-1$ and $0$. The probabilities are
$$
\begin{aligned}
p_a &= P(Y_i=-1) = P(X_i <a),\\
\quad p_b &= P(Y_i=1) = P(X_i > b),\\
\quad  p_{ab} &= P(Y_i=0) = P(X_i \in [a, b]) = 1-p_a-p_b. 
\end{aligned}
$$
We now use the specific form of the signal given in (\ref{eq:model}) 
and write
$$
\begin{aligned}
p_a &= P (X_i < a) = P (\theta + \epsilon_i < a)
= P(\epsilon_i < a - \theta)\\
&= F(a - \theta), \\
p_b &= P (X_i > b)  = P (\theta + \epsilon_i >b)
= P(\epsilon_i > b - \theta)\\
&= 1-F(b-\theta).  
\end{aligned}
$$
Since we assume that $F$ is known and invertible, we can identify $\theta$ 
by solving the two equations with respect to $\theta$, which gives
\be \label{eq:theta}
\theta = a - F^{-1}(p_a), \quad
\theta = b - F^{-1}(1- p_b).
\ee


The above formulas for $\theta$ 
involve the probabilities $p_a$ and $p_b$, which can be estimated with the (efficient) 
maximum likelihood estimator, 
$$
\hat{p}_a = \frac{1}{n}\sum_{i=1}^n \One\{X_i<a\} \quad \mbox{and} \quad
\hat{p}_b = \frac{1}{n}\sum_{i=1}^n \One\{X_i>b\}.
$$
Plugging these into (\ref{eq:theta}) yields two estimators for
$\theta$,
\be
\label{thhat}
\hat \theta_a = a - F^{-1}(\hat p_a)  \quad \mbox{and} \quad
\hat \theta_b = b - F^{-1}(1-\hat p_b).
\ee
Since $\hat \theta_a$ and $\hat \theta_b$ are continuous functions of
the consistent estimators $\hat p_a$ and $\hat p_b$, they 
are consistent estimators for $\theta$.

The estimator $\hat p_b$ (and analogously $\hat p_a$) multiplied by $n$ has a binomial distribution,
$$
n \hat p_b = \sum_{i = 1}^n \One\{X_i >b\} \sim {\cal B}in(n,p_b).
$$
An application of the central limit theorem and the delta method yield the approximate normal distribution of the estimators $\hat p_a$ and $\hat p_b$ and, in particular, for the two estimators of $\theta$, namely
\be
\label{AsNormtheta}
\hat \theta_a \approx 
{\cal{N}}\left(\theta, \frac{F(a - \theta) [1-F(a-\theta)]}{n f^2(a-\theta)}\right),
 \nonumber \\
\hat \theta_b \approx 
{\cal{N}}\left(\theta, \frac{F(b - \theta) [1-F(b-\theta)]}{n f^2(b-\theta)}\right).
\ee
The $f$ in the above formulas denotes the density of the error distribution $F$.
A more detailed derivation of these results is given in Section \ref{Appendix2} of the Appendix; see also Greenwood et al. \cite{greenwood1999}.

\subsection{Estimation of $\theta$ using weighted averages}\label{sec:weightedave}
The (approximate) variances of $\hat \theta_a$ and  $\hat \theta_b$, say $v_a$ and $v_n$, are provided in equation (\ref{AsNormtheta}). A well-established approach to combine two estimators for the same parameter is to use the (estimated) variances to construct weights, so that the better estimator (with a smaller variance) gets more weight. Using this method to combine $\hat \theta_a$ and $\hat \theta_b$ given in (\ref{thhat}) leads to
\be 
\label{eqn:estimator}
\hat \theta =  \left( \frac{\hat{v}_a  }
{\hat{v}_a + \hat{v}_b } \right) \hat{\theta_b} + \left( \frac{ \hat{v}_b }
{\hat{v}_a + \hat{v}_b } \right) \hat{\theta_a},
\ee
where $\hat{v}_a$ and $\hat{v}_b$ are empirical counterparts of $v_a$ and $v_b$,
\be \label{varab}
\hat v_a=\frac{\hat p_a(1-\hat p_a)}{n f^2(F^{-1}(\hat p_a))}, \quad
\hat v_b=\frac{\hat p_b(1-\hat p_b)}{n f^2(F^{-1}(\hat p_b))}.
\ee
This estimator is straightforward to implement since $f$ and $F$ are known and $\hat p_a$ and $\hat p_b$ are simply relative frequencies. In our simulations, we will work with a refined version of (\ref{eqn:estimator}) which uses more flexible kernel estimators for estimating $p_a$ and $p_b$ (see Section \ref{sec:NW} for more details). 

Another more complex estimator that combines $\hat \theta_a$ and $\hat \theta_b$ using optimal (minimum variance) weights is discussed in Section \ref{Appendix1} of the Appendix.

\subsection{Non-parametric estimators for $p_b$ and $p_a$}\label{sec:NW}
Up to here we have explained an estimation method for threshold data assuming that the underlying signal is constant. According to Greenwood et al. \cite{greenwood1999} and M{\"u}ller \cite{muller2000a}, it is also possible to recover non-constant smooth signals from these data if the noisy signal is characterized by a non-parametric regression model with independent errors. 

We now explain a procedure for estimating a non-constant probability function $p_b(\cdot)$. As seen earlier, the same method can be used to estimate $p_a$. 

The noisy signal at a time point $t_i$ is expressed as  
\be \label{eqn:nonconstant}
X(t_i) = \theta(t_i) + \epsilon(t_i), \quad i = 1,2, \cdots, n.
\ee 
Then the exceedance probability at the time point $t_i$ is  
\ba 
p_b(t_i) = P[X(t_i) > b]  = P[\theta(t_i) + \epsilon(t_i) > b]. 
\ea

Given that the distribution of $\epsilon(t_i)$, say $F$ (or $F_{t_i}$), has an inverse, then $\theta(t_i) = b - F^{-1}(1 - p_b(t_i))$. As suggested by M{\"u}ller in \cite{muller2000a}, the Nadaraya-Watson (NW) estimator is suitable for estimating $p_b(t_i)$ and $p_a(t_i)$:
$$ \displaystyle
\hat{p}_b(t_i) = \frac{\sum_{j = 1}^n \frac{1}{w} K\left(\frac{t_i - t_j}{w}\right)\One(X(t_i)>b)}{ \sum_{j = 1}^n \frac{1}{w} K\left(\frac{t_i - t_j}{w}\right) }
\quad \textrm{and} $$
$$
\hat{p}_a(t_i) = \frac{\sum_{j = 1}^n \frac{1}{w} K\left(\frac{t_i - t_j}{w}\right)\One(X(t_i)<a )}{ \sum_{j = 1}^n \frac{1}{w} K\left(\frac{t_i - t_j}{w}\right) },
$$
where $w$ is the bandwidth. 

Analogously to (\ref{thhat}), we obtain estimators for $\theta$ by plugging
in $\hat p_a$ and $\hat p_b$. The two resulting estimators for the signal at the time point $t_i$ are
\be \label{eqn:weighted_a_b}
\hat \theta_b(t_i) = b- F^{-1}(1 - \hat p_b(t_i)), \quad
\hat \theta_a(t_i) = a- F^{-1}(\hat p_a(t_i)). 
\ee
These estimators can be combined as discussed in equation (\ref{varab}) in section \ref{sec:weightedave}, now with weights that involve local variance estimators $\hat v_b(t_i)$
and $\hat v_a(t_i)$ instead of $\hat v_b$ and $\hat v_a$, i.e.\
\begin{eqnarray} 
\label{eqn:NPestimator}
\hat \theta(t_i) &=&  \left( \frac{\hat{v}_a(t_i)  }
{\hat{v}_a(t_i) + \hat{v}_b(t_i) } \right) \hat{\theta}_b(t_i) 
\nonumber \\
&&
+ \left( \frac{ \hat{v}_b(t_i) }
{\hat{v}_a(t_i) + \hat{v}_b(t_i) } \right) \hat{\theta}_a(t_i),
\end{eqnarray}
with
$$
\hat v_a(t_i)=\frac{\hat p_a(t_i)(1-\hat p_a(t_i))}{n f^2(F^{-1}(\hat p_a(t_i)))}, \quad
\hat v_b=\frac{\hat p_b((t_i)1-\hat p_b)(t_i)}{n f^2(F^{-1}(\hat p_b(t_i)))}.
$$


\subsection{Enhanced signal detection using wavelets}\label{SR_WR_Relation}
Consider the non-constant signal $\theta(t_i)$, where $i=1, \cdots, n$, described in (\ref{eqn:nonconstant}). Mathematically, the discrete wavelet transform (DWT) of the signal can be given in matrix form as follows
\begin{equation}\label{eq-12}
  \uw{d} = W\theta,
\end{equation}
where $W$ is an orthogonal matrix of size ${n \times n}$ (i.e., \(W^TW = I\)). The elements in $W$ are determined by selecting a particular wavelet basis, such as Haar, Daubechies, or Symmlet. Although $n$ can be arbitrary, it is usually selected to be a power of two, i.e., $n = 2^J$, $J \in \mathbb{Z}^+$ for the ease of calculating the DWT. 

The DWT of the signal is obtained by performing a series of convolutions that involve a wavelet-specific low-pass filter (scaling function) and its mirror counterpart high-pass filter (wavelet function). These convolutions using the two filters followed by downsampling (i.e., decimation operation that keeps every second element of the convolution) result in a multiresolution representation of the signal. The transformed signal consists of a smoothing approximation ($c-$coefficients) and a hierarchy of detail coefficients ($d_{jk}$) at different resolutions, given by a scale index $j$, and different locations $k$ within the same $j$th resolution. As the convolutions with these two filters are repeated until a desired decomposition level $j = J_0$ is reached ($1 \leq J_0 \leq J-1$), the approximation coefficients represent progressively smoother versions of the signal as high-frequency content is removed. Thus, the transformed signal $\uw{d}$ has the following structure 

\begin{figure*}[!t]
        \centering
            \includegraphics[width= 2\columnwidth]{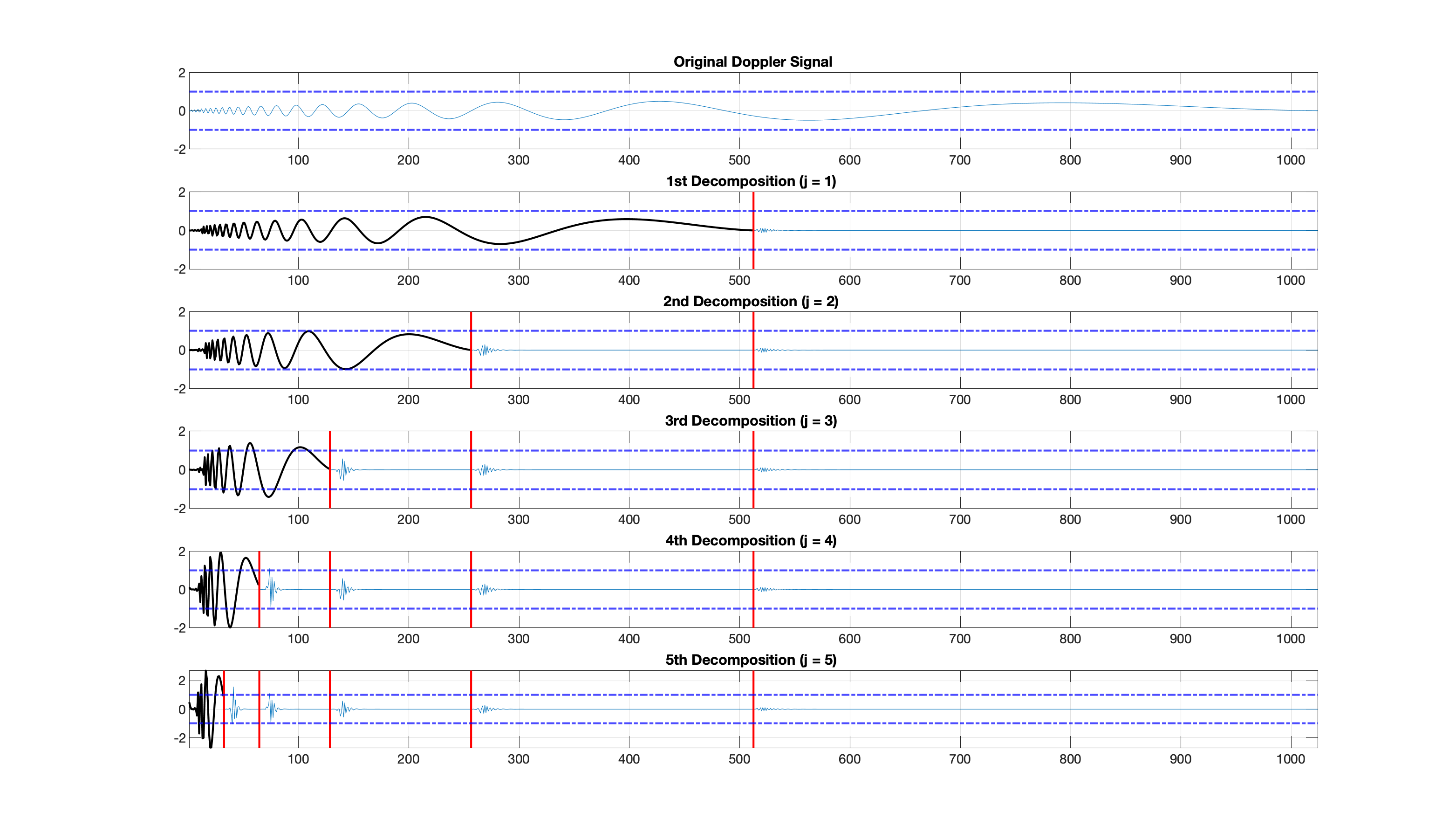}
\caption{  Wavelet decomposition of a Doppler signal of length \(1024\) (\(J = \log_2(1024) = 10\)), illustrating the amplification of the wavelet-transformed signal beyond the two thresholds \(a = -1\) and \(b = 1\). The decomposition progresses across levels \(j = 1\) to \(5\). At each level, the signal is separated into smoothing (black) and detail (blue) components. The red lines indicate the boundaries where the wavelet decomposition occurs, visually marking the separation between components at different levels of resolution. This figure demonstrates how the signal structure evolves with finer decompositions. } 
    \label{fig:SR_wt}
\end{figure*}

\begin{equation}\label{eq-12a}
  \uw{d} = (\uw{c}_{J_0}, \uw{d}_{J_0}, \dots, \uw{d}_{J-2}, \uw{d}_{J-1}),
\end{equation}
where $\uw{c}_{J_0}$ is a vector of coefficients corresponding to a smooth trend in signal, and $\uw{d}_{j}$ are detail coefficients at different resolutions $j$ where $J_0 \leq j \leq J-1.$ 

In addition, multiplication of (\ref{eq-12}) by the transformed wavelet matrix $W^T$, the original signal can be reconstructed as follows

\be \label{eqn:iwt}
\theta = W^T\uw{d}. 
\ee

Notably, through this recursive decomposition, the approximation coefficients continually enhance the signal’s low-frequency representation, resulting in a smoother approximation at each level. This progressive smoothing amplifies both the smooth and detail coefficients of the signal. As a result, the likelihood that the signal exceeds specific thresholds is higher in the multiscale domain than in the original data domain.  

For example, Figure \ref{fig:SR_wt} demonstrates the wavelet decomposition process applied to a Doppler signal of length \(1024\), with a maximum decomposition level of \(J = \log_2(1024) = 10\). The decomposition is illustrated for levels \(j = 1\) to \(5\), showing how the signal is progressively analyzed at different scales. The top-left panel displays the original Doppler signal, while the subsequent panels depict the decomposed signal at each level. At each decomposition level, the signal is divided into two components: a smoothing component (plotted in black) that represents the low-frequency content (e.g., overall trend) and a detail component (plotted in blue) that isolates the high-frequency content, representing finer variations. The red vertical lines indicate the boundaries where the wavelet decompositions occur, marking the separation between the smoothing and detail components. As the decomposition level increases, the smoothing component becomes coarser, while the detail component captures increasingly finer-scale variations in the signal. This figure highlights the multiscale nature of wavelet decomposition, effectively demonstrating how the signal's structure evolves across different resolutions, showcasing the potential of weak signal detectability with less noise compared to the data domain. (Section \ref{sec:Discussion} provides a detailed description of the amplification of the wavelet-transformed signal, exceeding the thresholds.)
 
The procedure for signal recovery in the multiscale domain follows a similar approach to the one described in Sections \ref{sec:thhat} and \ref{sec:NW}. The main steps are as follows:  

\begin{enumerate}  
    \item {\bf Adding noise to the wavelet-transformed signal:}  
    Noise is introduced to the transformed signal \(\uw{d}\), defined in (\ref{eq-12}), as:  
    \be  
    \uw{d}_e(t_i) = \uw{d}(t_i) + e(t_i), \quad i = 1, \cdots, n,  
    \ee  
    where \(e(t_i)\) are independent and identically distributed (i.i.d.) random variables with a mean of zero and variance \(\sigma^2\), following a known invertible distribution function \(F\).  

    Since \(\uw{d}\) is a fixed vector, \(\uw{d}_e(t_i)\) follows the same distribution \(F\), but with a mean of \(\uw{d}\) and a variance of \(\sigma^2\). 

    \item {\bf Computing exceedance probabilities:}  
    Similar to the method for non-constant signals in (\ref{eqn:nonconstant}), we compute the probabilities that the noisy transformed signal \(\uw{d}_e\) exceeds predefined thresholds at time \(t_i\):  
    $$  
    p_{(\uw{d})b}(t_i) = P[\uw{d}_e(t_i) > b]~\textrm{and}~p_{(\uw{d})a}(t_i) = P[\uw{d}_e(t_i) < a].  
    $$  
    The Nadaraya-Watson (NW) estimator, described in Section \ref{sec:NW}, is used to estimate these exceedance probabilities, denoted as \( \hat{p}_{(\uw{d})a}(t_i)\) and \( \hat{p}_{(\uw{d})b}(t_i)\).  

    \item {\bf Estimating the transformed signal \(\uw{d}\):}  
    Using the exceedance probability estimators, we estimate the transformed signal \(\uw{d}\) at each time point \(t_i\) as follows:  
    \be \label{eqn:weighted_a_b_wt}  
        \hat{\uw{d}_{b}}(t_i) = b - F^{-1}[1 - \hat{p}_{(\uw{d})b}(t_i)], \nonumber \\ 
        \hat{\uw{d}_{a}}(t_i) = a - F^{-1}[\hat{p}_{(\uw{d})a}(t_i)].  
    \ee  

    As in the non-wavelet case described in Section \ref{sec:thhat}, \(\hat{\uw{d}_{b}}(t_i)\) and \(\hat{\uw{d}_{a}}(t_i)\) follow a similar distribution, as given in (\ref{AsNormtheta}). Their variances, according to (\ref{varab}), are:  

    \be \label{varab_wt}  
    V(\hat{\uw{d}_a}) = \frac{\hat{p}_{(\uw d)a}(1 - \hat{p}_{(\uw d)a})}{n f^2(F^{-1}(\hat{p}_{(\uw d)a}))}, \nonumber \\  
    V(\hat{\uw{d}_b}) = \frac{\hat{p}_{(\uw d)b}(1 - \hat{p}_{(\uw d)b})}{n f^2(F^{-1}(\hat{p}_{(\uw d)b}))}.  
    \ee  

    These variance estimates are then used to refine the estimation of the transformed signal \(\uw{d}\):  

    \be \label{eqn:wt_estimator}  
    \hat{\uw{d}} = \Big[ \frac{V(\hat{\uw{d}_a})}  
    {V(\hat{\uw{d}_a}) + V(\hat{\uw{d}_b})} \Big] \hat{\uw{d}_b} +  
    \Big[ \frac{V(\hat{\uw{d}_b})}  
    {V(\hat{\uw{d}_a}) + V(\hat{\uw{d}_b})} \Big] \hat{\uw{d}_a}.  
    \ee  

    \item {\bf Recovering the original signal \(\theta\):}  
    Finally, as described in (\ref{eqn:iwt}), applying the inverse wavelet transform  to \(\hat{\uw{d}}\) yields an estimate of the original signal:  

    \ba  
    \hat{\theta} = W^T \hat{\uw{d}}.  
    \ea  
\end{enumerate}  

\subsection{Goodness of signal recovery and Fisher information}
In non-parametric regression theory, i.e.\ when estimating an unknown non-constant function, the average mean squared error (AMSE) is generally used as a measure to evaluate the goodness of the estimation,
\be 
AMSE =  \frac{1}{n} \sum_{i =1}^n{\mathbb E [\hat \theta(t_i) - \theta(t_i)]^2}. 
\ee 

\noindent This study uses the AMSE as a criterion for selecting an optimal bandwidth and noise level, and for evaluating the goodness of the recovered signal. Note that the AMSE reduces to the mean squared error $\mathbb E[\hat \theta-\theta]^2$ if a constant signal $\theta$ is estimated. More information about the use of the Nadaraya-Watson estimator in SR can be found in \cite{muller2000a}, where an estimator for a non-constant sub-threshold signal (with one threshold) is proposed. 

In Section 5 of \cite{muller2000a} an approximation of the local AMSE, 
\be \label{AMSE}
AMSE(t) = \mathbb E [\hat \theta(t) - \theta(t)]^2
\ee
is presented, which consists of two terms. The first term corresponds to the Fisher information, the inverse asymptotic variance of an efficient estimator for $\theta(t)$, and is responsible for the typical SR behavior: the information increases with increasing $\sigma$ until an optimal noise level is reached and then decreases since too much noise drowns out the signal.

The Fisher Information (FI) is therefore an important tool to demonstrate SR and to measure how much information about a constant parameter $\theta$ (or a local parameter $\theta(t)$ if the signal is non-constant) is retained by thresholding. It is extensively studied in Greenwood et al. \cite{greenwood1999} for the case of a constant sub-threshold signal.

To study FI more closely, we briefly review some formulas from \cite{greenwood1999}. 

Assume the noisy signal is observed directly,
$X_i = \theta + \epsilon_i$. Then the FI is
$$
I = \int m(x)^2 f(x) \, dx 
$$
with $m = f'/f$ the score function. If the signal is sub-threshold with
one threshold $b$, the FI has the form
$$
I_\theta^b = 
\frac{\Big(\int_{b-\theta}^\infty m(x) f(x) \, dx\Big)^2}
{F(b-\theta)[1-F(b-\theta)]}.
$$
The proportion of information $I_\theta^b/I$ retained by thresholding
is therefore $I_\theta^b/I = R(b-\theta)$
with
$$
R(u) =
\frac{\Big(\int_{u}^\infty m(x) f(x) \, dx\Big)^2}
{F(u)[1-F(u)] \Big(\int_{u}^\infty m(x) f(x) \, dx\Big)^2}.
$$
If $F=\Phi$ and $f=\phi$ are the cdf and pdf of the standard normal 
distribution, then $m(x) = x$ and 
$$
R(u) =\frac{\phi(u)^2} {\Phi(u)[1-\Phi(u)]}.
$$
If the errors $\epsilon_i$ have a distribution function 
$F_\sigma = F(x/\sigma)$,
where $F$ has variance $1$ and $\sigma$ is a scale parameter, the scale 
function becomes $m((b-\theta)/\sigma)/\sigma$ and the proportion of
information is
$$
\frac{I_{\theta \sigma}^b}{I_\sigma} = R\Big(\frac{b-\theta}{\sigma}\Big).
$$
If the noise is normally distributed with mean $0$ and variance $\sigma^2$, 
i.e.\ $F_\sigma(x) = \Phi(x/\sigma)$, this becomes
$$
\frac{I_{\theta \sigma}^b}{I_\sigma} = 
\frac{\phi \Big(\frac{b-\theta}\sigma\Big)^2} 
{\Phi \Big(\frac{b-\theta}\sigma\Big)
\Big[1-\Phi \Big(\frac{b-\theta}\sigma\Big)\Big]}.
$$
We can now consider our case where only time points are observed when the noisy signal is above the threshold
$b$ ($Y_i=1$), below a second threshold $a$ ($Y_i=-1$) or between the two
thresholds ($Y_i=0$). 
As in \cite{greenwood1999}, Section III, where a scenario with a sub-threshold signal 
{\em below  several} thresholds is discussed, one obtains 
$$  \tiny 
I_{\theta \sigma}^{ab} = \frac1{\sigma^2}  \left( 
\frac{ \phi \Big(\frac{a-\theta}\sigma \Big)^2 }
     { \Phi \Big(\frac{a-\theta}\sigma \Big) }
+
\frac{\Big[\phi\Big(\frac{b-\theta}\sigma\Big) 
       - \phi\Big(\frac{a-\theta}\sigma \Big)\Big]^2 }
     { \Phi \Big(\frac{b-\theta}\sigma \Big)
       - \Phi \Big(\frac{a-\theta}\sigma \Big) }
+
\frac{ \phi \Big(\frac{b-\theta}\sigma \Big)^2 }
     { 1-\Phi \Big(\frac{b-\theta}\sigma \Big) }
\right).
$$

Since $I = 1/\sigma^2$, the proportion of information 
$I_{\theta \sigma}^{ab}/I$ retained by the
$Y_i$'s is  $\sigma^2 I_{\theta \sigma}^{ab}$.

\begin{figure}[!t]
        \centering
            \includegraphics[width=\columnwidth]{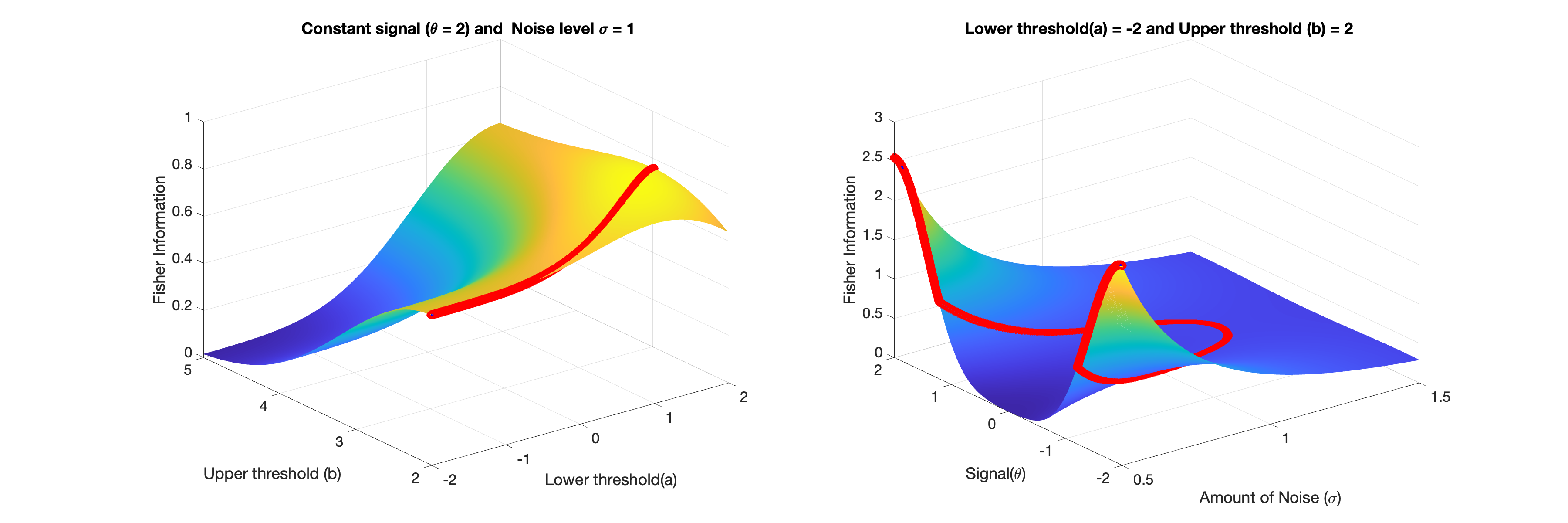}
             \caption{Visualization of Fisher Information (FI): Sensitivity to thresholds, signal amplitude, and noise levels. The left panel illustrates the dependence of FI on lower and upper thresholds (\(a\) and \(b\)) for a constant signal (\(\theta = 2\)) under a fixed noise level (\(\sigma = 1\)). The right panel depicts how FI varies with signal amplitude (\(\theta\)) and noise level (\(\sigma\)), given fixed thresholds (\(a = -2\) and \(b = 2\)). The red curves highlight the maximum FI, revealing the influence of parameter tuning on signal detectability.}
             \label{fig:FI}
\end{figure}

Figure \ref{fig:FI} presents a visualization of Fisher Information in relation to various parameters. The left panel shows how Fisher Information varies with the upper threshold (\(b\)) and lower threshold (\(a\)) for a constant signal (\(\theta = 2\)) under a fixed noise level (\(\sigma = 1\)). The red curve highlights the path of the maximum FI achieved for each combination of \(a\) and \(b\). The highest FI \(0.8098\) was achieved when the signal is closer to the thresholds \(a = 1.3884\) and \(b = 2.6116\). The right panel depicts the variation of FI with the signal amplitude (\(\theta\)) and noise level (\(\sigma\)), given fixed thresholds (lower \(a = -2\) and upper \(b = 2\)). The red curve emphasizes critical regions where FI peaks with respect to the noise level. The highest FI, 1.5465 was achieved when the signal is equal to the threshold ( i.e., \(\theta = a = -2\) and \(\theta= b =2\). Notably, if the signal is 0 (neither close to \(a\) nor to \(b\)) it needs the most noise to shift the noisy signal above \(b\) or below \(a\). Overall, these plots demonstrate the sensitivity of Fisher Information to key parameters, aiding in understanding the trade-offs and optimal configurations for signal detection. 


\section{Performance evaluation}\label{sec:Simulation}
This section presents the weak signal detection performance of the double-threshold detector through simulations, comparing it to sub- and sup-threshold detectors. A periodic sub-threshold signal is said to be weak if the amplitude is twice as low as the threshold of the detector. The proposed detector is tested on (simulated and benchmark) 1D and 2D signals, with recovery assessed in both data and multiscale domains. All performance evaluations are conducted in MATLAB, and the evaluation software is available on {\it GitHub}.

\subsection{Methodology}\label{sec:sim}

\subsubsection{Signal detection based on simulated 1D and 2D signals}\label{sec:sim-sim} 

\begin{figure}[!t]
        \centering
            \includegraphics[width= 1\columnwidth]{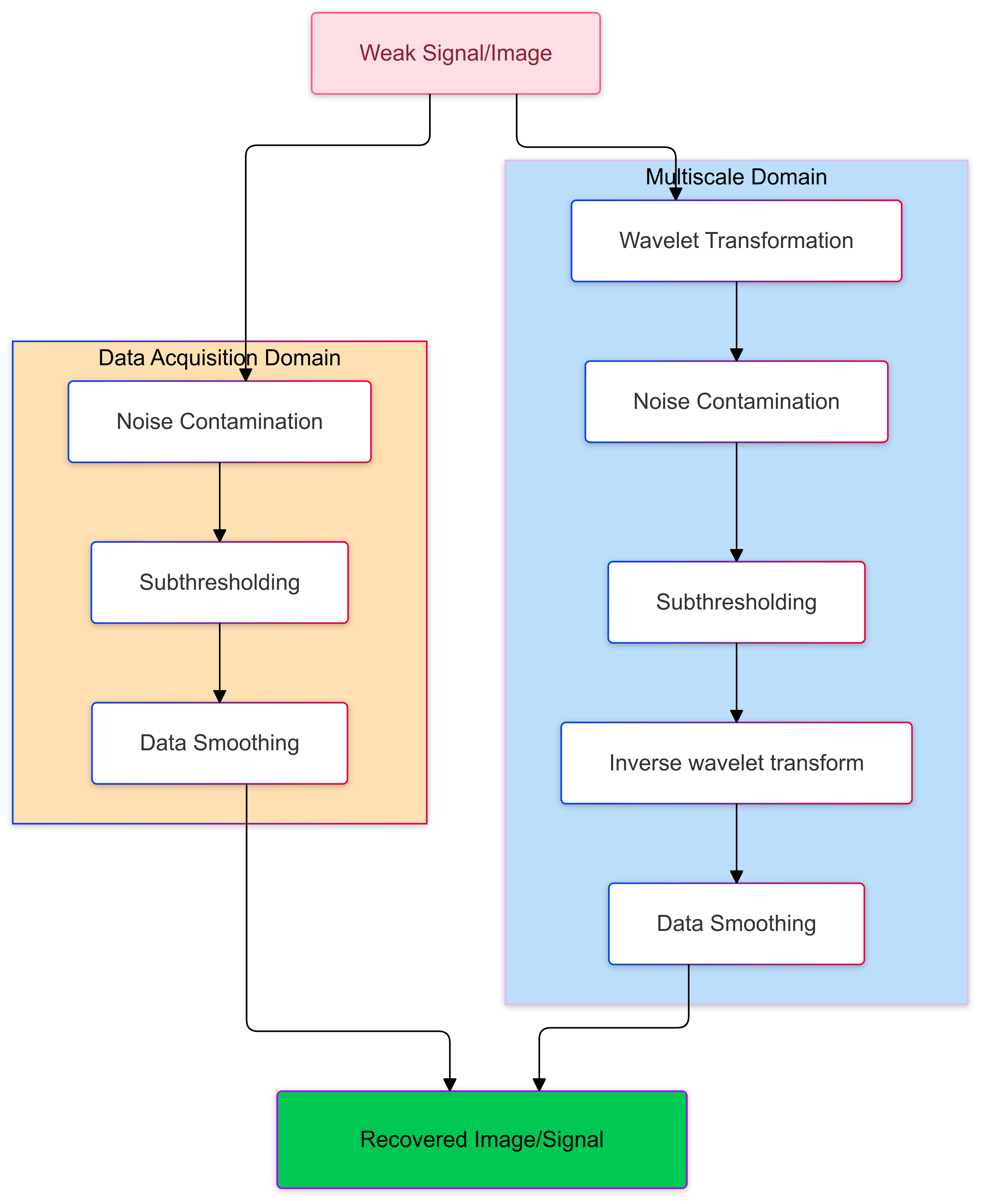}
             \caption{An overview of the procedures used for the detection of (1D and 2D) weak signals in the data acquisition and multiscale(frequency/wavelet) domains. In the original data domain, thresholding is applied to the original signal, while in the multiscale domain, thresholding is applied to the wavelet-transformed data. }
             \label{fig:flow}
    \end{figure}

\paragraph{1D signals} We simulate a sine wave ($\theta = \sin(x)$, where $0 \leq x \leq 8\pi$) with 1,024 data points (i.e., signal length $n = 1,024$). The thresholds for the detectors are $a = 2\min(\theta) = -2$ and $b= 2 \max(\theta) =2$. These values ensure that the thresholds are at least twice as large as the actual signal values. In the signal recovery process, since this is a time-varying signal, we use the NW estimator-based non-parametric approach, given in equation \eqref{eqn:NPestimator}), to compute the threshold exceedance probability values. 

\paragraph{2D signals}  An image ($\theta= \sin(l)\cos(m)$, where $l,m \in [\pi, 8\pi]$) of size $512 \times 512$ is generated. The thresholds are $a = 2 \min(y) = -2$ and $b = 2\max(y) =2$. When computing the recovery signal, as this is not a time-varying signal, we use the parametric approach presented in Section \ref{sec:weightedave} to estimate exceedance probability values. 

Figure \ref{fig:flow} gives an overview of the main steps in the weak signal detection procedure in the data acquisition and the multiscale domain. Here are the details.

    \begin{enumerate}
        \item {\bf Signal recovery in the data (acquisition) domain:} Following the discussion in Section \ref{sec:Problem}, the main steps of the signal recovery procedure involve noise contamination, thresholding, exceedance probability estimation, and signal recovery calculations. The optimal noise level ($\sigma$) for accurate signal recovery is determined by varying $\sigma$ for a range of levels and assessing the recovery performance using the Average Mean Squared Error (AMSE) given in (\ref{AMSE}). 
        
        In the AMSE calculation process, the recovery signal estimation process is repeated 100 times at each noise level. This is followed by computing the average of the squared difference between the actual signal (i.e., \(\theta_{n \times 1}\)) and the mean of the recovered 100 signals (i.e., \(\hat \theta_{n\times 1}\)). The noise level that minimizes the AMSE is selected as the optimal amount of noise required for the best signal recovery.  
        
        For time-varying signals, i.e.\ 1D signals considered in this study, in addition to searching for an optimal noise level $\sigma$, we also obtain the optimal NW bandwidth $w$ that ensures the best signal recovery performance. The AMSE evaluation process described above is employed for a range of $w$ values. The noise level and bandwidth values that minimize the AMSE are selected as the optimal amount of noise and bandwidth for the best signal detection. 

        \item {\bf Signal recovery in the multiscale (wavelet/frequency) domain:} The experimental settings and steps are similar to those used in the previous data domain scenario. The main difference is that noise contamination, thresholding, and exceedance probability estimation are performed on the wavelet-transformed signal as described in section \ref{SR_WR_Relation}. The wavelet filter used here is the {\it Symmlet} filter of size 8 ({\it Symmlet-8}). The reason for using this filter over the other shorter filters, such as {\it Daubechies-2 (Haar)} and {\it Daubechies-4}, is that it offers a better balance of smoothness, near-symmetry, and locality, for accurate signal approximation and cleaner reconstructions. 
    \end{enumerate}

The results of our evaluation are presented in sections (\ref{SR-time}) and (\ref{SR_WT}). 
    
\subsubsection{Signal detection based on benchmark signals}\label{sec:sim_bm}
Using the signal detection procedures just described, we also compare the signal detection performance of the double-threshold detector in the data and wavelet domains using commonly used benchmark signals. This includes six 1D signals (\textit{Wave, Doppler, Time-shifted sine, Blip, Angles, Parabolas}) of length 1024 and four 2D images: \textit{Barbara, Cameraman, Mandrill}, and \textit{Peppers} of size \(512 \times 512\). The criteria used to select thresholds (\(a, b\)), noise level (\(\sigma\)), and kernel bandwidth (\(w\)) are similar to those used in simulation studies in Section \ref{sec:sim-sim}. The results are provided in Section \ref{SR-benchmark}.
 
\subsection{Results}\label{sec:results}
\subsubsection{Signal detection in the data domain}\label{SR-time}

\begin{figure*}[t!]
\centering
\begin{subfigure}{ 1\columnwidth}
  \centering
  \includegraphics[width=1\columnwidth]{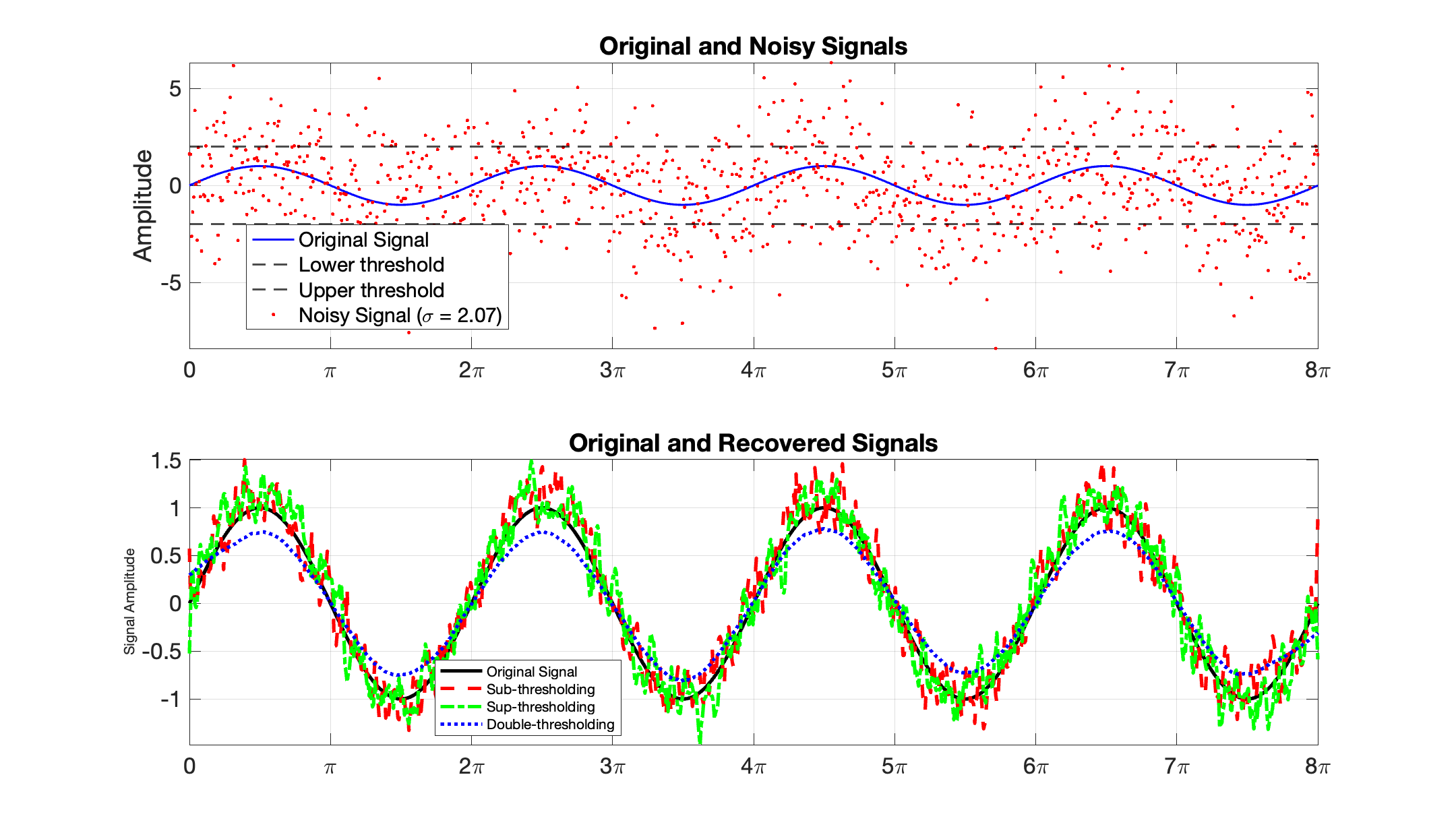}
  \caption{}
  \label{data_1d}
\end{subfigure}
\begin{subfigure}{1\columnwidth}
  \centering
  \includegraphics[width= 1\columnwidth]{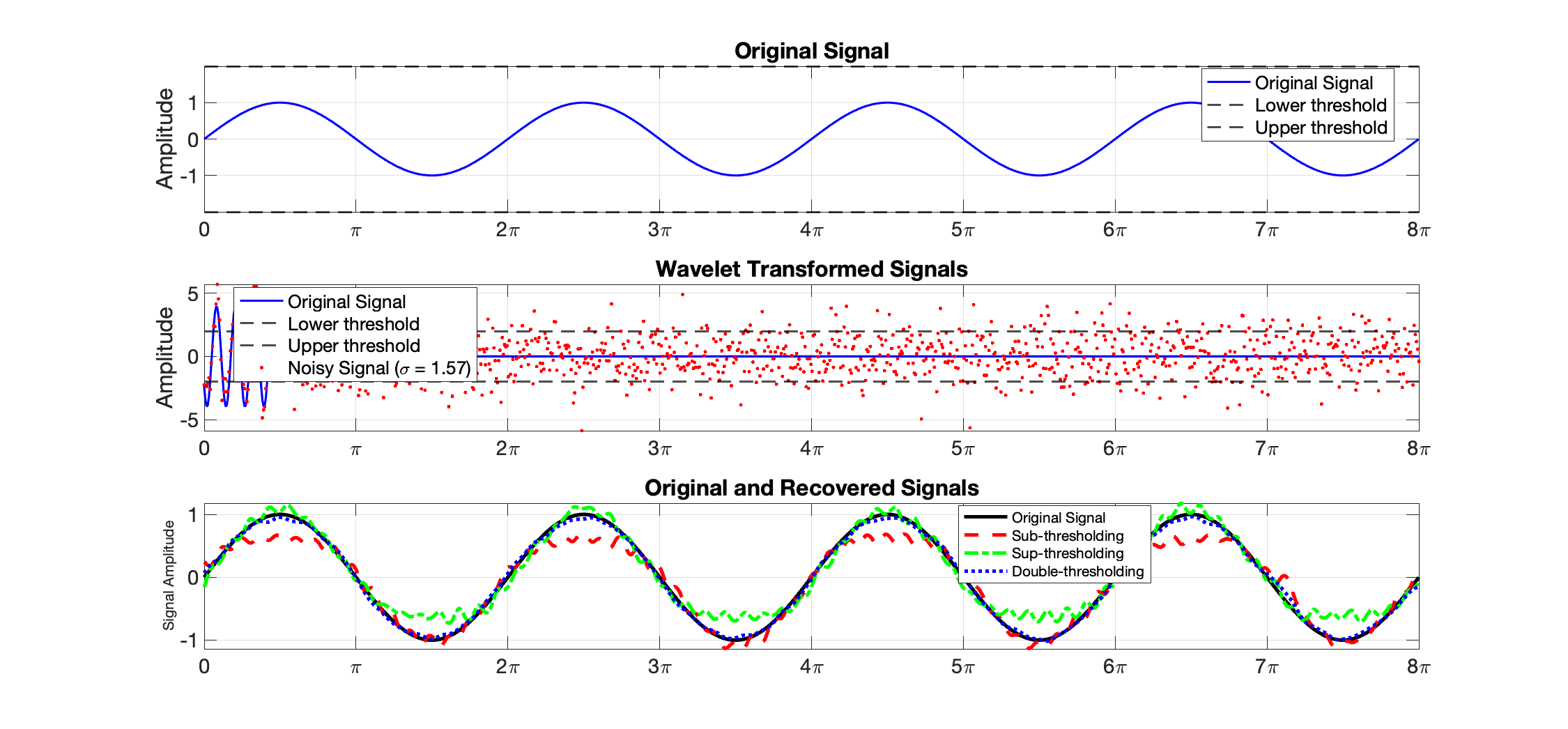}
  \caption{}
  \label{fig1D_wt}
\end{subfigure}
\caption{One-dimensional signal recovery performance with original data domain (a) and frequency/multiscale domain (b). The graph on (a) shows the original signals as well as thresholds and a noise sample. The second graph shows the recovered signals from the sub-, the sup-, and the double-threshold detectors. In the frequency domain, (b), noise contamination is conducted in the wavelet domain (second graph) and recovery from the three detectors is shown on the third graph.}
\label{fig:sin1}
\end{figure*}

We begin with a discussion of the 1D signal detection performance. As illustrated in Figure \ref{data_1d}, the original \emph{sine} wave remains entirely within the thresholds \(a = -2\) and \(b = 2\), rendering it undetectable by sub-, sup-, and double-threshold detectors. However, introducing zero-mean normal noise with standard deviation \(\sigma = 2\) (red dots in the first graph of Figure \ref{data_1d}) causes some values to exceed these thresholds, enabling signal recovery through exceedance probabilities (\(p_a, p_b\)).  

Table \ref{tab:1} summarizes the optimal noise level \(\sigma\) and corresponding AMSE values when $\sigma$ ranges from 1 to 5 and $w$ from $0.001$ to $50$. The double-threshold detector achieves the lowest AMSE (\(0.0325\)) for $\sigma = 2.0713$, outperforming the sub-threshold detector (AMSE = \(0.0373\), \(\sigma = 2.4248\)) and the sup-threshold detector (AMSE = \(0.0375\), \(\sigma = 2.2733\)). The optimal bandwidths for the sub-, sup-, and double-threshold detectors are \(0.0122\), \(0.0172\), and \(0.1000\), respectively.  

Using these optimal parameters, the second graph in Figure \ref{data_1d} displays the recovered signals alongside the original signal. The double-threshold detector produces a recovered signal (blue) that closely follows the original signal (black), in contrast to the sub- and sup-threshold detectors (green and red lines). 

\begin{figure*}[!t]
\centering   
\includegraphics[width=1\textwidth]{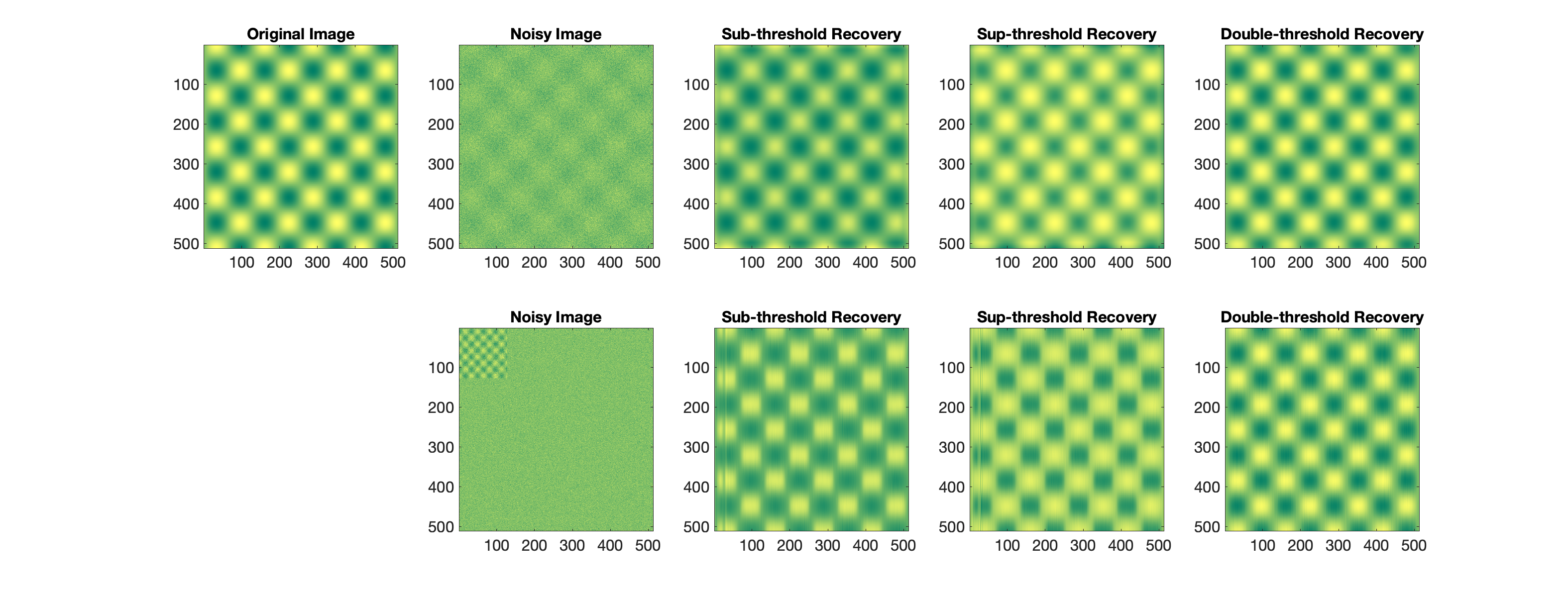}
\caption{ Two-dimensional signal recovery performance is analyzed for the original data (first row) and in the frequency domain (second row) using the image $\theta = \sin{l}\cos{m}$ of size $512 \times 512$. The first column presents the original, while the second column contains the noisy version of the original and the wavelet-transformed image. The third, fourth, and fifth columns show the recovered signals using sub-, sup-, and double-threshold detectors, respectively.}
\label{fig:2d}
\end{figure*}

Following the approach used for 1D signals (except for bandwidth selection), the analysis was extended to the 2D images \( y(l,m) = \sin(l)\cos(m) \). The first row of Figure \ref{fig:2d} shows the original and the noisy image, as well as the recovered images using sub-, sup-, and double-threshold detectors with thresholds \( a = -2 \) and \( b = 2 \).  
AMSE was analyzed for noise levels $\sigma \in [0.25, 2]$, with optimal values summarized in Table \ref{tab:1}. Similar to the 1D case, the double-threshold detector achieves superior performance (AMSE = 0.0012) compared to sub- and sup-threshold detectors (AMSE = 0.3187 and 0.3184). The optimal noise level for the double-threshold detector (\(\sigma = 1.5395\)) is slightly higher than that for the sub- and sup-threshold detectors (\(\sigma = 1.3553\)).  

\subsubsection{Signal recovery in the multiscale domain}\label{SR_WT}
In the 1D signal analysis, we applied the wavelet transform using the \textit{Symmlet-8} filter with three decomposition levels (\(J_0 = 4\)). Table \ref{tab:1} summarizes the optimal noise level \(\sigma\) and the corresponding AMSE values obtained by evaluating \(\sigma \in [1, 4]\) and \(w \in [0.0001, .1]\). The optimal bandwidths for the sub-, sup-, and double-threshold detectors are \(0.0102\), \(0.0213\), and \(0.0213\), respectively. 

Consistent with the results from the data domain, the double-threshold detector achieves the lowest AMSE (\(0.0018\)), closely following the original signal, in contrast to the sub- and the sup-threshold detectors (AMSE $0.0319$ and $0.0291$, respectively). It requires less noise (\(\sigma = 1.5663\)) for optimal performance than the corresponding approach in the data domain. Figure \ref{fig1D_wt} presents the recovered signals using the optimal \(\sigma\) and \(w\) values.  

For 2D signal recovery, as shown in Figure \ref{fig:2d} (second row), the double-threshold detector reconstructs the image more accurately than the sub- and sup-threshold detectors in the multiscale domain. Table \ref{tab:1} reports an AMSE of \(0.0017\) for the double-threshold detector (\(\sigma = 1.3353\)), whereas the sub- and sup-threshold detectors yield AMSE values \(0.212\) and \(0.0218\) with \(\sigma = 1.4474\).  

\begin{table*}[t!]
    \centering
    \begin{tabular}{lccccccc}
    \hline
&  & \multicolumn{3}{c}{Data Acquisition Domain} & \multicolumn{3}{c}{\textbf{Multiscale  Domain}}\\
    \hline
&  & \multicolumn{3}{c}{Threshold Detector} & \multicolumn{3}{c}{Threshold Detector}  \\
&  & Sub & Sup & \textbf{Double} & Sub & Sup & \textbf{Double} \\
    \cline{2-8}
  \multirow{2}{*}{1D}  & Noise Level($\sigma$) & 2.4248  & 2.2733 & {\bf 2.0713} & 2.2228 & 2.1723 & {\bf 1.5663} \\
    \cline{2-8}
    & AMSE   & 0.0373  & 0.0375 & \textbf{0.0325} & 0.0319 & 0.0291 & {\bf 0.0018} \\ 
        \cline{2-8}
  \multirow{ 2}{*}{2D}  & Noise Level($\sigma$) & 1.3553 & 1.3553 & \textbf{1.5395} & 1.4474 & 1.4474 & {\bf 1.3553} \\
    \cline{2-8} 
        & AMSE & 0.3187 & 0.3184 & \textbf{0.0012} & 0.0212 & 0.0218 &  {\bf 0.0017}\\ \hline
    \end{tabular}
    \caption{Signal recovery performance (AMSE) of the threshold detectors with the optimal noise level ($\sigma$) selected from analyzing the 1D sine wave and 2D image generated using \(\sin(l)\cos(m)\) in the original and multiscale domains.}
    \label{tab:1}
\end{table*}

\subsubsection{Benchmark signal recovery}\label{SR-benchmark}
This section illustrates for several benchmark signals the recovery performance of the double-threshold detector in the data and in the frequency domains. We use the 1D and 2D recovery procedures described earlier. \textbf{These benchmark signals are designed to replicate a diverse range of real-world signals commonly encountered in applications. Widely recognized in signal processing research, they serve as standard test cases for evaluating methods. Specifically, they enable a meaningful assessment of the proposed method's performance in practical scenarios.}

\begin{figure*}[!t]
    \centering
    \includegraphics[width= 1\textwidth]{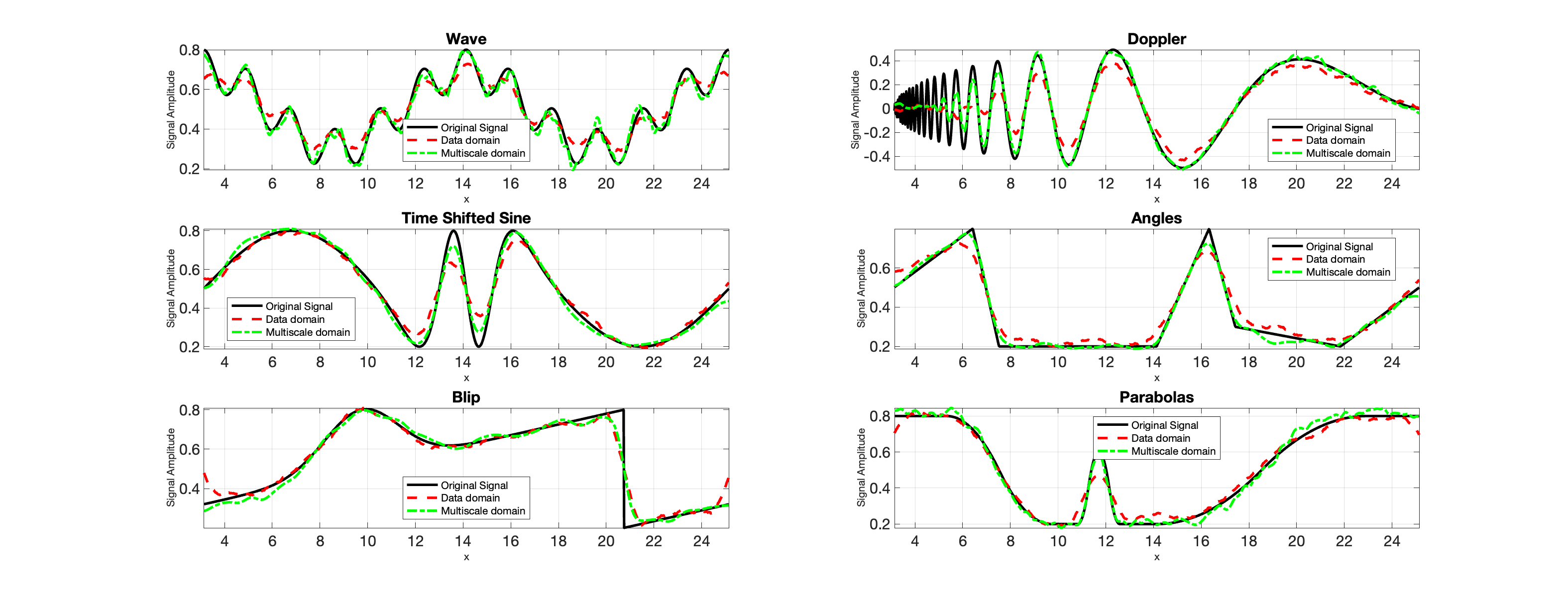}
    \caption{Signal recovery visualization: double-threshold detector performance across 1D test signals in data and multiscale domains.}
    \label{test_1D_signals}
\end{figure*}

\begin{figure*}[!t]
    \centering
    \includegraphics[width= 1\textwidth]{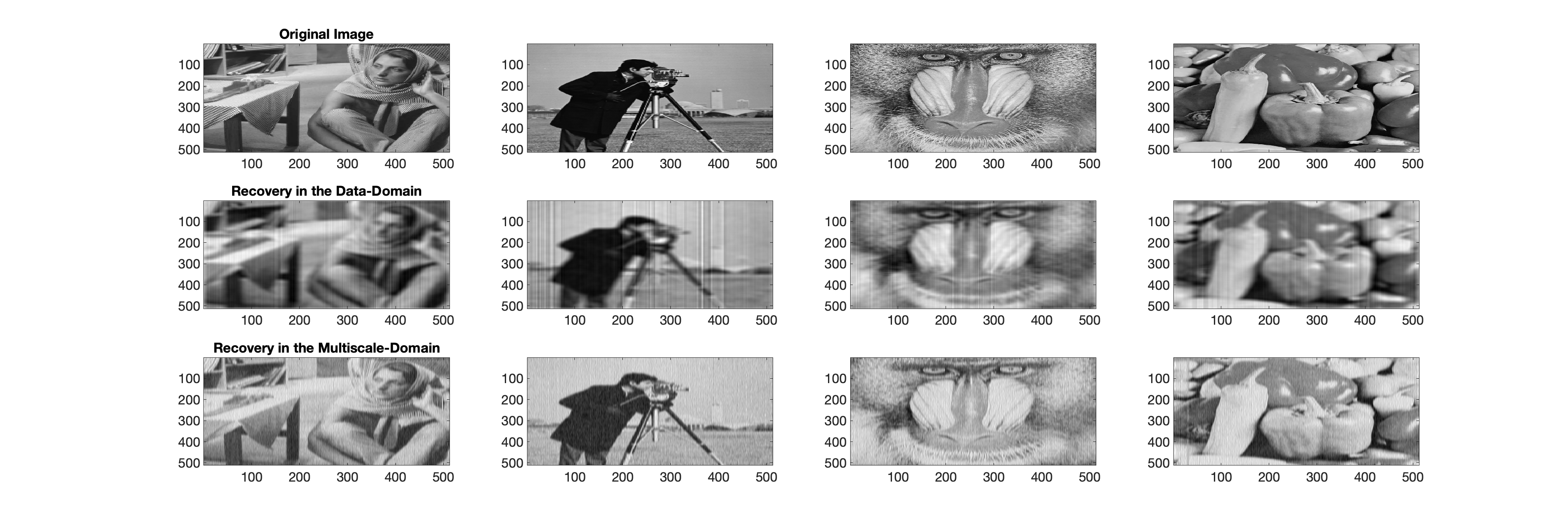}
    \caption{Signal recovery visualization: double-threshold detector performance across 2D test images in data and multiscale domains.}
    \label{fig_4s}
\end{figure*}

Figure \ref{test_1D_signals} shows the original and recovered 1D signals, while Figure \ref{fig_4s} shows the 2D signals. Signals recovered with the double-threshold detector closely follow the originals in both domains. Recoveries in the data domain exhibit greater deviations compared to the multiscale domain. Table \ref{tab:2} summarizes the optimal noise levels and AMSE values. The results confirm that the multiscale domain consistently achieves lower noise levels and AMSE values across all benchmark signals.  

For 1D signals, noise levels in the multiscale domain are significantly reduced compared to the data acquisition domain. Consider, for example, the \textit{Wave} signal. Here the noise decreases from \(1.3273\) to \(0.8172\). In the \textit{Doppler} signal example, we have a decline from \(1.4293\) to \(0.7151\). AMSE values also show substantial reductions, particularly for the \textit{Angles} signal (\(0.0014\) to \(0.0003\)) and the \textit{Parabolas} signal (\(0.0012\) to \(0.0005\)).  

For 2D signals, noise reduction is less pronounced, but still evident. The \textit{Barbara}, \textit{Cameraman}, \textit{Mandrill}, and \textit{Peppers} signals exhibit lower noise levels in the multiscale domain (\(0.8944\) or \(1.1156\)) compared to the data acquisition domain (\(1.1156\) in all cases). AMSE values also improve, particularly for the \textit{Cameraman} signal, where AMSE decreases from \(0.0150\) to \(0.0062\), demonstrating a significant accuracy enhancement.  

Overall, the multiscale domain effectively reduces noise and improves AMSE across both 1D and 2D signals, with particularly strong noise suppression in the 1D cases. These results highlight the superior performance of the double-threshold detector in the multiscale domain.

\begin{table*}[t!]
    \centering
    \begin{tabular}{llcccc}
    \hline
     & Test Signal    & \multicolumn{2}{c}{Data Acquisition Domain} & \multicolumn{2}{c}{Multiscale  Domain} \\
         \hline
    &                 & Noise Level ($\sigma$) & AMSE & Noise Level ($\sigma$) & AMSE  \\
    \hline
\multirow{6}{*}{1D} & Wave & 1.3273   & 0.0020 &  0.8172  &  0.0006 \\
                    & Doppler &   1.4293 & 0.0124   & 0.7151 & 0.0033 \\
                    & Time-shifted sine & 1.1232 & 0.0019 & 0.6131 & 0.0007 \\
                    & Angles & 1.2252 & 0.0014  & 0.7151 & 0.0003 \\
                    & Blip   & 1.1232 & 0.0021  & 0.7151 & 0.0015 \\
                    & Parabolas & 1.1232 & 0.0012 & 0.6131 & 0.0005 \\
                    \hline
\multirow{4}{*}{2D} & Barbara & 1.1156  & 0.0087 & 0.8944 & 0.0082 \\  
                    & Cameraman & 1.1156  & 0.0150 & 1.1156 & 0.0062 \\ 
                    & Mandrill & 1.1156  & 0.0075 & 0.8944 & 0.0141 \\ 
                    & Peppers & 1.1156  & 0.0082 & 0.8944 & 0.0067 \\ 
                    \hline
     \end{tabular}
    \caption{ Performance of the double-thresholding detector: comparative noise levels ($\sigma$) and AMSE for 1D and 2D test signal recoveries in data acquisition and multiscale domains.}
    \label{tab:2}
\end{table*}
\section{Discussion}\label{sec:Discussion}
The exploration of weak signal detection performance shows that the multi-threshold detector makes a superior contribution to improve SR performance compared to the standard SR methods that are based on single-threshold detectors. This contribution is further enhanced when these detectors are employed in the multiscale domain of the signal. 
This section discusses the importance of multi-threshold detectors, weak signal detection in the multiscale domain, including possible directions in which the study could be explored further.  

\subsection{Motivation of using wavelets}\label{subsec:motivation}
Consider a communication setup in which Alice transmits a weak signal to Bob. However, Bob's observation system has a hard threshold: he can only register values of the signal that lie outside a fixed band. Signals within this band are invisible to him, effectively resulting in a saturated or clipped observation model. To aid in the transmission, an intermediary, Charlie, is allowed to preprocess the signal before Bob receives it. Charlie is constrained: he cannot alter the signal's mean or variance, that is, he cannot add a bias, and cannot amplify the signal. However, he is permitted to apply an invertible, lossless transformation. Charlie will inform Bob about the exact procedure used, unless Bob already knows the method and has agreed upon in advance.
 
Charlie chooses to apply an orthogonal discrete wavelet transform (DWT) to the signal, decomposing it up to  $L$
levels. Orthogonal wavelets are energy-preserving and invertible, satisfying Charlie’s constraints. Bob, in turn, receives the transformed signal (in the wavelet domain), and seeks to recover Alice’s original signal using stochastic resonance techniques.
 
Here is where wavelets play a critical role: in the DWT, much of the signal’s low-frequency energy is concentrated in the approximation (scaling) coefficients. As the decomposition proceeds across levels, the scaling coefficients are updated recursively, and their magnitude increases approximately by a factor of $\sqrt{2}$ per level (assuming normalized wavelets). As visualized in Figure \ref{fig:SR_wt}, this effectively brings the signal energy closer to Bob’s threshold boundaries, even though the overall variance remains unchanged. As a result, the transformed signal in the wavelet domain is more likely to cross the observation threshold, even with a smaller amount of added noise.
 
This shift is crucial: in conventional SR, if the signal is far from the threshold, large noise is needed to induce threshold crossings; however,  large noise also corrupts the signal and deteriorates recovery accuracy. The wavelet transform allows for reduced noise amplitude, enhancing the precision and reliability of the SR-based reconstruction. Once Bob applies SR in the wavelet domain and collects sufficient threshold crossings, he performs the inverse wavelet transform, using Charlie’s metadata, and reconstructs the original signal in the time domain with greater fidelity.
 
In summary, wavelets assist SR-based recovery in thresholded systems by redistributing energy across scale, concentrating it in components that are more likely to exceed observation bounds under limited noise, and thereby enabling more efficient and accurate signal reconstruction.

\subsection{The importance of using multiple threshold detectors}
The illustrative examples presented in Section \ref{sec:Simulation} reflect that the double-threshold detector performs better than the single-threshold detectors in terms of their AMSE presented in Table \ref{tab:1} and \ref{tab:2}. These findings confirm the idea that multi-thresholding improves the signal detection proposed by Greenwood et. al. in \cite{greenwood1999} based on constant signals. This improvement is because the double-threshold detector can capture more information than single-threshold detectors. More specifically, as described in Section \ref{sec:Problem} the double-threshold detectors used in this study account for the exceedance events from both sub- and sup-thresholding methods (or uses both probabilities $p_a$ and $p_b$) in the signal recovery process. Whereas, the single-threshold detectors rely on exceedance events from sub- or sup-thresholding (or use probability value either $p_a$ or $p_b$). This results in obtaining more information about signal underlying dynamics to obtain a better estimator for the signal (or $\theta$ in \ref{eqn:nonconstant}).

\subsection{Advantages of performing SR in the multiscale domain}
As evident in Table \ref{tab:1} and Table \ref{tab:2}, incorporating multiscale signal analysis methods could further enhance weak signal detection. That is, sub-thresholding in the multiscale (wavelet) domain offers significant improvements in weak signal recovery due to the synergistic benefits of both techniques. The wavelet transform's multiresolution analysis allows signals to be decomposed into different frequency bands, enabling SR to enhance weak signals at the relevant scales while reducing noise in irrelevant ones. This selective enhancement is further supported by the wavelet domain's spatial and frequency localization, which allows for more precise amplification of weak signals that are transient or non-stationary in nature. By leveraging noise to improve signal detection, SR becomes more effective in the wavelet domain as noise is distributed across frequency bands, allowing for adaptive thresholding and selective noise reduction at specific scales. This helps improve the signal-to-noise ratio (SNR), as the noise in high-frequency bands can be suppressed while the weak signal is amplified in the appropriate frequency range. Additionally, the non-linear system dynamics inherent in SR are optimized by the wavelet transform’s ability to separate signal components, further enhancing the detection of weak signals. Consequently, this integration of SR and wavelet transforms offers a robust framework for improving weak signal recovery, particularly in noisy environments, by exploiting noise in a constructive manner and allowing better signal isolation in the time-frequency domain.

\begin{figure}[t!]
        \centering
            \includegraphics[width= \columnwidth]{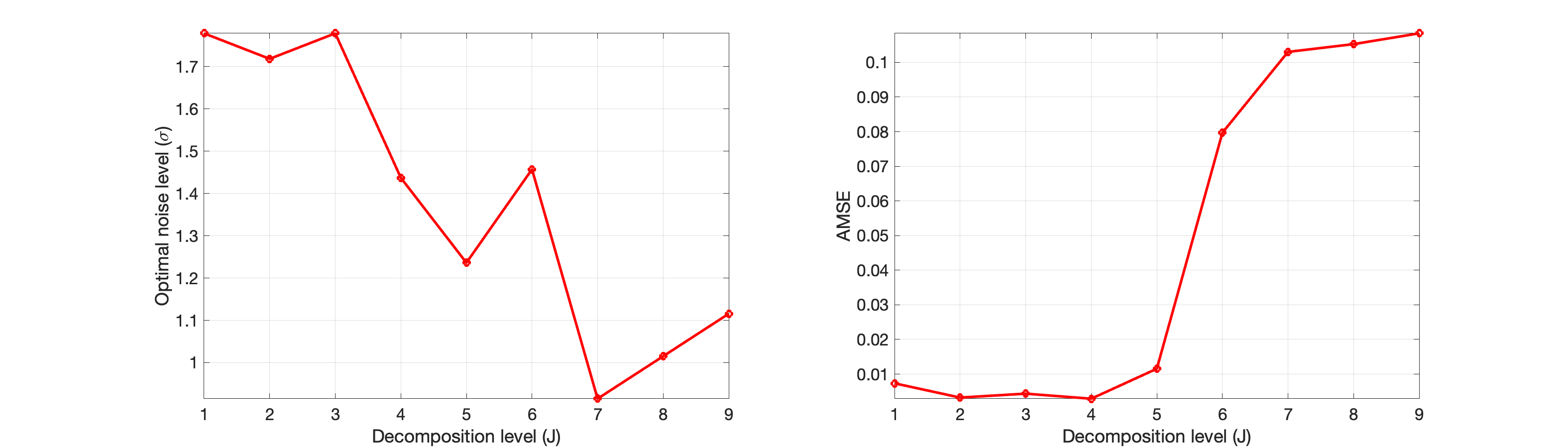}
            \caption{Signal recovery performance with multiresolution level. The sine wave used in 1D signal recovery is used to assess recovery performance for different wavelet decomposition levels $j$.}
            \label{fig1D_AMSE_J}
    \end{figure}

As described in Section \ref{SR_WR_Relation}, higher wavelet decomposition levels increase the likelihood of the transformed signal exceeding specific thresholds. This, in turn, reduces the amount of noise ($\sigma$) required to detect weak signals. To explore the relationship between noise level ($\sigma$) and wavelet decomposition level ($J_0$), the multiscale 1D signal recovery process was repeated at decomposition levels. The results, presented in Figure \ref{fig1D_AMSE_J}, demonstrate that the optimal noise level ($\sigma$) decreases as the wavelet decomposition level increases. Consequently, the Average Mean Squared Error (AMSE) also decreases. These findings indicate that higher decomposition levels enhance signal recovery performance. 

However, as seen in AMSE vs decomposition level $j$ in Figure \ref{fig1D_AMSE_J}, excessive separation of smoothing and high-frequency properties of the signal does not improve SR performance. That is because excessive wavelet decomposition can degrade signal properties by causing loss of important features, introducing numerical artifacts, reducing signal-to-noise ratio, and leading to over-smoothing. It increases computational complexity and may introduce reconstruction errors or aliasing. Optimal decomposition levels should be carefully chosen based on the signal's frequency content, application needs, and empirical testing to balance detail preservation and noise reduction while avoiding unnecessary deterioration.

\subsection{Importance of the type of noise}
Since noise plays a key role in SR, it is crucial to carefully consider the suitability of its source (i.e., the noise distribution $F$ in \eqref{eq:model}) in relation to the signal dynamics. The normal distribution has traditionally been a common choice for generating noise due to its wide applicability. However, in many scenarios, the true noise distribution is unknown. In such cases, an effective approach is to explore various noise distributions, allowing for a comparative analysis of signal recovery under different conditions. Notably, even when the noise distribution is incorrectly specified, the underlying structure of a signal can still be inferred from an estimated version, particularly after applying a scale transformation. Therefore, analyzing the signal detection performance of the double-threshold detector across different noise distributions is a worthwhile consideration. However, this remains a potential extension for future research, as the primary focus of this study is to evaluate the effectiveness of multi-threshold detectors in the frequency domain.


\section{Concluding remarks}\label{sec:Conclusion}

This paper introduces a novel double-threshold detector and evaluates its performance in weak signal detection. The proposed detector is a weighted average of sub- and sup-threshold detectors. Its effectiveness is examined in both the data and the multiscale (frequency) domains using 1D and 2D simulated and benchmark signals.  

Our findings demonstrate that the double-threshold system, grounded in solid theoretical principles, significantly improves weak signal detection compared to single-threshold systems. Experimental results indicate that its performance is further enhanced when applied in the multiscale domain, achieving substantial noise reduction. These results highlight the effectiveness of performing stochastic resonance in the multiscale domain, specifically by integrating stochastic resonance with wavelet transform, offering practical approaches for detecting weak signals in dynamic and noisy environments.  

However, certain limitations must be acknowledged. Firstly, the experimental evaluations in this study considered only normal noise for weak signal recovery. Future work should investigate the impact of different noise distributions and optimize parameters accordingly. Secondly, the proposed detector is a variance-based estimator. As discussed in Section \ref{Appendix1}, incorporating covariance between the sub- and sup-threshold detectors in the derivation of the double-threshold detector could further enhance its performance.

\section*{Acknowledgments}
This research was supported in part by the H.O. Hartley endowment at Texas A\&M University. The study sponsor had no involvement in the study design, collection, analysis and interpretation of the data; in the writing of the manuscript; and in the decision to submit the manuscript for publication.

\section*{Appendix}
\begin{appendices}
\subsection{An advanced method for estimating $\theta$} \label{Appendix1}
In this article we have used a straightforward variance-based weighted estimator; see equation (\ref{eqn:estimator}). Here we describe another weighted estimator that is similar, but more complex. We explain the approach for the case of a constant signal $\theta$. The generalization to the case with a non-constant signal $\theta(t)$ is straightforward: simply replace the constant weights $w$ by local weights $w(t)$.

Assume $\hat \theta_a$ and $\hat \theta_b$ are two estimators with finite variances $v_a$ and $v_b$ for some some parameter $\theta$ (not necessarily our $\hat \theta_a$ and $\hat \theta_b$ from Section \ref{sec:weightedave}). Consider the weighted average 
$$
w \hat \theta_a + (1-w) \hat \theta_b, \quad w \in [0, 1].
$$
Optimal weights (with minimum variance) $\omega$ and $1-\omega$ can be obtained by differentiating the variance with respect to $\omega$ and setting it equal to zero. The variance is
$$
\omega^2 v_a + (1-\omega)^2 v_b + 2 w(1-w) C,
$$
where $C$ denotes the covariance between $\hat \theta_a$ and $\hat \theta_b$, i.e.\ $C=cov(\hat \theta_a, \hat \theta_b)$. We differentiate the variance with respect to $\omega$ and set the derivative equal to zero and obtain
$$
\begin{aligned}
&2 \omega v_a - 2 (1-\omega) v_b + 2 (1-2w) C = 0\\
&\iff \omega v_a - v_b + \omega v_b + C - 2wC = 0\\ 
& \iff  \omega [v_s + v_b - 2C] = v_b -C,
\end{aligned}
$$
which yields the optimal weights
\begin{equation} \label{optimalw}
\omega^* = \frac{v_b -C}{v_a + v_b - 2C}, \qquad
1-\omega^* = \frac{v_a -C}{v_a + v_b - 2C}.
\end{equation}
Our estimator from equation (\ref{eqn:estimator}) is a special case with $C=0$.

The covariance can be estimated with bootstrap, or numerically, using the fact that we have a multinomial distribution. The covariance is
$$
\begin{aligned}
C & = Cov\left(a-F^{-1}(\hat p_a), b-F^{-1}(1- \hat p_b)\right) \\ 
  & = Cov\left(F^{-1}(\hat p_a),F^{-1}(1- \hat p_b)\right) \\ 
  & = \mathbb{E}[F^{-1}(\hat p_a) \cdot F^{-1}(1- \hat p_b)] -  \mathbb E[F^{-1}(\hat p_a)] \cdot \mathbb E[F^{-1}(1- \hat p_b)],
\end{aligned}
$$
but the expectations can be computed numerically, using the fact that the counts $(n \hat p_a, n \hat p_{ab}, n \hat p_b)$ have a multinomial distribution. Then plug in $\hat p_a$ and $\hat p_b$ for the unknown probabilities $p_a$ and $p_b$. 

We have, for example,
$$ \tiny \displaystyle 
\mathbb E[F^{-1}(\hat p_a) \cdot F^{-1}(1- \hat p_b)] = \mathbb E \left[F^{-1} \left(\frac{n \hat p_a}{n}\right) \cdot F^{-1}\left(\frac{n-n \hat p_b}{n}\right) \right].
$$
The explicit formula for the expectation, with $\hat p_a$ and $\hat p_b$  in place of $p_a$ and $p_b$ is

\ba \displaystyle
\mathop{\sum_{x_a=0}^n \sum_{x_b=0}^n}_{x_a+x_b \le n} 
F^{-1}\left(\frac{x_a}{n}\right) \cdot F^{-1}\left(\frac{n-x_b}{n}\right)
\frac{n!}{x_a!x_b!(n-x_a-x_b)!} \\
.\hat p_a^{x_a} \hat p_b^{x_b} (1-\hat p_a - \hat p_b)^{n-x_a-x_b}. 
\ea
The estimators for the other two expectations can be calculated using the binomial distribution.


\subsection{Proof of equation~\eqref{AsNormtheta}}\label{Appendix2}
The two formulas for $\theta$ in (\ref{eq:theta}),
$$
\theta = a - F^{-1}(p_a), \quad
\theta = b - F^{-1}(1- p_b),
$$
suggest estimating $\theta$ by plugging in
$$
\hat{p}_a = \frac{1}{n}\sum_{i=1}^n \One\{X_i<a\} \quad \mbox{and} \quad
\hat{p}_b = \frac{1}{n}\sum_{i=1}^n \One\{X_i>b\},
$$
which yields two estimators for $\theta$,
$$
\hat \theta_a = a - F^{-1}(\hat p_a)  \quad \mbox{and} \quad
\hat \theta_b = b - F^{-1}(1-\hat p_b).
$$
Since $\hat \theta_a$ and $\hat \theta_b$ are continuous functions of
the consistent estimators $\hat p_a$ and $\hat p_b$, they are
consistent for $\theta$ by the continuous mapping theorem.

The estimator $\hat p_b$ multiplied with $n$ is a sum of independent Bernoulli variables with the same parameter $p_b = 1-F(b-\theta)$ and thus has a binomial distribution,
$$
n \hat p_b = \sum_{i = 1}^n \One\{X_i >b\} \sim {\cal B}in(n,p_b),
$$
with expected value $np_b$ and variance $np_b(1-p_b)$.

By the central limit theorem, $(n\hat p_b - np_b)/\sqrt{np_b(1-p_b)}$
is asymptotically standard normally distributed, that is,
$$
\sqrt{n} (\hat{p}_b - p_b ) \sim {\cal AN}(0, p_b (1-p_b)).
$$
An application of the delta method yields
\be
\label{anthb}
 \sqrt{n} (\hat{\theta_b} - \theta) \sim {\cal AN}\left(0, \frac{F(b - \theta) (1-F(b-\theta)}{f^2(b-\theta)} \right),
 \ee
where $f$ is the density of the error distribution $F$. 

To see this, let $g(\hat{p}_b) = \hat{\theta} = b - F^{-1}(1-\hat{p}_b)$ with $\hat p_b$ approximately normally distributed,
$\hat p_b \approx {\cal N}(p_b, p_b(1-p_b)/n)$. Then, by the delta method,
$$\hat \theta_b = g(\hat{p}_b) 
\approx {\cal{N}}\left(g(p_b), \frac{(g'(p_b))^2 p_b(1-p_b)}{n}\right)
$$
with 
$$
\begin{aligned}
g(p_b) &=\theta, \quad  p_b(1-p_b) = (1-F(b-\theta))F(b-\theta),\\
g'(p_b) &= - \frac d{dp_b} F^{-1}(1-p_b) \\
&= - \frac 1{F'(F^{-1}(1-p_b))}\\
&= - \frac 1{f(F^{-1}[F(b-\theta)])}\\
&= - \frac 1{f(b-\theta)}.
\end{aligned}
$$
Inserting this into the above $\hat \theta_b$ yields
$$
\hat \theta_b \approx 
{\cal{N}}\left(\theta, \frac{p_b(1-p_b)}{n f^2(F^{-1}(1-p_b))}\right) $$
$$=
{\cal{N}}\left(\theta, \frac{F(b - \theta) [1-F(b-\theta)]}{n f^2(b-\theta)}\right),
$$
i.e.\ the asymptotic statement given in (\ref{anthb}).

\noindent Similarly, for ${\hat \theta_a} = a - F^{-1}(\hat{p}_a)$, we obtain analogous statements, in particular
$$ 
\begin{aligned}
\sqrt{n} (\hat{p}_a - p_a ) &\sim {\cal AN}\left(0, p_a (1-p_a)\right), \\
\hat \theta_a &\approx 
{\cal{N}}\left(\theta, \frac{p_a(1-p_a)}{n f^2(F^{-1}(p_a))}\right) \\
&={\cal{N}}\left(\theta, \frac{F(a - \theta) [1-F(a-\theta)]}{n f^2(a-\theta)}\right).
\end{aligned}
$$ 
\end{appendices}

\bibliographystyle{IEEEtran}
\bibliography{sample}

\end{document}